\documentclass[longauth,letter,online]{aa}
\usepackage{epstopdf}
\usepackage{graphicx}
\usepackage{rotating}
\usepackage{txfonts}
\usepackage{color}
\usepackage{natbib}
\bibpunct{(}{)}{;}{a}{}{,}
\usepackage{ulem}
\usepackage[colorlinks = true,
            allcolors  = blue]{hyperref}

\makeatletter
\renewcommand*\aa@pageof{, page \thepage{} of \hyperref[LastPage]{\pageref{LastPage}}}
\makeatother
\usepackage{multirow}
\usepackage{bm}
\defcitealias{xxlpaperI}{Paper~I}
\defcitealias{xxlpaperII}{Paper~II}
\defcitealias{xxlpaperIV}{Paper~IV}
\defcitealias{xxlpaperXIX}{Paper~XIX}
\defcitealias{xxlpaperXX}{Paper~XX}

\doi{\href{https:doi.org/10.1051/0004-6361/201834022}{https:doi.org/10.1051/0004-6361/201834022}}

\begin{document}
\idline{A\&A, ???, L? (2018)}
\topics{The XXL Survey: second series}
\AANum{L?}
\date{Received 4 August 2018 / Accepted 11 September 2018}

\newcommand{\dd}{deg$^{2}$}
\newcommand{\flux}{$\rm erg \, s^{-1} \, cm^{-2}$}
\newcommand{\LL}{$\Lambda$}

        \title{
                The XXL Survey
                \thanks{Based on observations obtained with {\it XMM-Newton}, an ESA science mission with
        instruments and contributions directly funded by ESA Member States and NASA.}
        }  
        \subtitle{XXV. Cosmological analysis of the C1 cluster number counts}

        \author{
        F.~Pacaud\inst{\protect\ref{AIfA}}
        \and M.~Pierre\inst{\protect\ref{CEAdap}} 
        \and J.-.B.~Melin\inst{\protect\ref{CEAdphp}}
        \and C.~Adami\inst{\protect\ref{LAM}}
        \and A.~E.~Evrard\inst{\protect\ref{UmichPhysics},\protect\ref{UmichAstro}}
        \and S.~Galli\inst{\protect\ref{IAP}}
        \and F.~Gastaldello\inst{\protect\ref{IASFmilano}}
        \and B.~J.~Maughan\inst{\protect\ref{Bristol}}
        \and M.~Sereno\inst{\protect\ref{ObsBologna},\protect\ref{UnivBologna}}
        \and S.~Alis\inst{\protect\ref{Istanb}}
        \and B.~Altieri\inst{\protect\ref{ESAC}}
        \and M.~Birkinshaw\inst{\protect\ref{Bristol}}
        \and L.~Chiappetti\inst{\protect\ref{IASFmilano}}
        \and L.~Faccioli\inst{\protect\ref{CEAdap}}
        \and P.~A.~Giles\inst{\protect\ref{Bristol}}
        \and C.~Horellou\inst{\protect\ref{Chalmers}}
        \and A.~Iovino\inst{\protect\ref{Brera}}
        \and E.~Koulouridis\inst{\protect\ref{CEAdap}}
        \and J.-P.~Le~F\`evre\inst{\protect\ref{CEAdedip}}
        \and C.~Lidman\inst{\protect\ref{ANU}}
        \and M.~Lieu\inst{\protect\ref{ESAC}}
        \and S.~Maurogordato\inst{\protect\ref{Lagrange}}
        \and L.~Moscardini\inst{\protect\ref{ObsBologna},\protect\ref{UnivBologna},\protect\ref{INFNbologna}}
        \and B.~M.~Poggianti\inst{\protect\ref{INAFpadova}}
        \and E.~Pompei\inst{\protect\ref{ESO}}
        \and T.~Sadibekova\inst{\protect\ref{CEAdap}}
        \and I.~Valtchanov\inst{\protect\ref{ESAC}}
        \and J.~P.~Willis\inst{\protect\ref{UVic}}
        }

        \institute{
                Argelander-Institut f\"ur Astronomie,
        University of Bonn,
        Auf dem H\"ugel 71,
        D-53121 Bonn,
        Germany\label{AIfA}\\
            \email{fpacaud@astro.uni-bonn.de}
                \and
                AIM, CEA, CNRS, Universit\'e Paris-Saclay, Universit\'e Paris Diderot, Sorbonne Paris Cité, F-91191 Gif-sur-Yvette, France\label{CEAdap}
                \and
                IRFU, CEA, Universit\'e Paris-Saclay, F-91191 Gif-sur-Yvette, France\label{CEAdphp}
                \and 
                Universit\'e Aix Marseille, CNRS, LAM, Laboratoire d'Astrophysique de Marseille, Marseille, France\label{LAM}
                \and
                Department of Physics and Michigan Center for Theoretical Physics, University of Michigan, Ann Arbor, MI 48109, USA\label{UmichPhysics}
                \and 
                Department of Astronomy, University of Michigan, 
                Ann Arbor, MI 48109, USA\label{UmichAstro}
                \and
                Institut d'Astrophysique de Paris (UMR7095: CNRS \& UPMC-Sorbonne Universities), F-75014, Paris, France\label{IAP}
                \and 
                INAF - IASF Milano, via Bassini 15, I-20133 Milano, Italy\label{IASFmilano}
                \and
                HH Wills Physics Laboratory, Tyndall Avenue, Bristol, BS8 1TL, UK\label{Bristol}
                \and 
                INAF-OAS Osservatorio di Astrofisica e Scienza dello Spazio di Bologna, via Gobetti 93/3, 40129, Bologna, Italy\label{ObsBologna}
                \and
                Dipartimento di Fisica e Astronomia, Alma Mater Studiorum Università di Bologna, via Gobetti 93/2, 40129 Bologna, Italy\label{UnivBologna}
                \and 
                Department of Astronomy and Space Sciences, Faculty of Science, 41 Istanbul University, 34119 Istanbul, Turkey\label{Istanb}
                \and 
                European Space Astronomy Centre (ESA/ESAC), Operations Department, Villanueva de la Can\~ada, Madrid, Spain\label{ESAC}
                \and 
                Chalmers University of Technology, Dept of Space, Earth and Environment, Onsala Space Observatory, SE-439 92 Onsala, Sweden\label{Chalmers}
                \and 
                INAF - Osservatorio Astronomico di Brera, Via Brera 28, 20122 Milano, via E. Bianchi 46, I-20121 Merate, Italy\label{Brera}
        \and 
                CEA Saclay, DRF/Irfu/DEDIP/LILAS, 91191 Gif-sur-Yvette Cedex, France\label{CEAdedip}
                \and 
                Research School of Astronomy and Astrophysics, Australian National University, Canberra, ACT 2611, Australia\label{ANU}
                \and 
                Laboratoire Lagrange, UMR 7293, Universit\'e de Nice 
                Sophia Antipolis, CNRS, Observatoire de la C\^ote d'Azur, F-06304 Nice, France\label{Lagrange}
        \and 
                INFN - Sezione di Bologna, viale Berti Pichat 6/2, 40127, Bologna, Italy\label{INFNbologna}
                \and
                INAF- Osservatorio astronomico di Padova, Vicolo Osservatorio 5, I-35122 Padova, Italy\label{INAFpadova}
                \and
                European Southern Observatory, Alonso de Cordova 3107, Vitacura, 19001 Casilla, Santiago 19, Chile\label{ESO}
        \and
        Department of Physics and Astronomy, University of Victoria, 3800 Finnerty Road, Victoria, BC V8P 1A1, Canada\label{UVic}
        }

  \abstract
  {We present an estimation of cosmological parameters with clusters of galaxies.}
  {We  constrain  the $\Omega_m$, $\sigma_8$, and $w$ parameters from  a stand-alone sample of X-ray clusters detected in the 50 deg$^2$ XMM-XXL survey with a well-defined selection function.}
  {We analyse the redshift distribution of a sample comprising 178 high S/N clusters out to a redshift of unity. The cluster sample scaling relations are determined in a self-consistent manner.}
  {In a lambda cold dark matter $(\Lambda$CDM) model, the cosmology favoured by the XXL clusters compares well with results derived from the Planck S-Z clusters for a totally different sample (mass/redshift range, selection biases, and scaling relations). However, with this preliminary sample and current mass calibration uncertainty, we find no inconsistency with the Planck CMB cosmology. If we relax the $w$ parameter, the Planck CMB uncertainties increase by a factor of  $\sim$10 and become comparable with those from XXL clusters. Combining the two probes allows us to put constraints on $\Omega_m=0.316\pm0.060$, $\sigma_8=0.814\pm0.054$, and $w=-1.02\pm0.20$.}
  {This first self-consistent cosmological analysis of a sample of serendipitous XMM clusters already provides interesting insights into the constraining power of the XXL survey.  
  Subsequent analysis will use a larger sample extending to lower confidence detections and include additional observable information, potentially improving posterior uncertainties by roughly a factor of 3.}

\titlerunning{Cosmological analysis of the XXL C1 galaxy cluster sample}
\authorrunning{F. Pacaud et al.}
\maketitle

\section{Introduction}
Recent observations of the cosmic microwave background (CMB) by the Planck mission 
have resulted in a new set of cosmological constraints with unprecedented precision 
\citep{planck2015XIII}. While these measurements still remain entirely consistent with 
the simplest six-parameter lambda cold dark matter $(\Lambda$CDM)  Universe, they also reveal inconsistencies
between the interpretation of the Cosmic Microwave Background data and several of 
the late time cosmological probes,
in particular a $>3\sigma$ tension with local measurements of the Hubble constant using Cepheids \citep[e.g.][]{riess18}, as well as a higher predicted amplitude of matter fluctuations in the 
late time Universe  compared to cosmic shear measurements \citep{joudaki17,hildebrandt17}\footnote{However, 
some other recent studies do not reproduce these inconsistencies, e.g. \citet{troxel18}}
or the observed number counts of galaxy clusters \citep{planck2015XXIV}. 

While part of these discrepancies could be accounted for by statistical fluctuations,
investigating their origin could also point to new physics beyond the basic 
$\Lambda$CDM model or reveal residual systematics that remain to be understood in the 
interpretation of the different probes.
For instance, while some work has pointed to a moderately high value for the neutrino mass 
($0.1 \lesssim \sum{m_\nu} \lesssim 0.5\,\mathrm{eV}$) as a plausible solution for the dearth of  
massive clusters in the local Universe \citep{planck2015XXIV,salvati18}, others invoke systematic 
uncertainty in the cluster mass scale estimate as the main route to softening the discrepancy \citep{vonderlinden14,israel15,sereno17}.
Indeed, while some recent results use a weak lensing mass calibration \citep[e.g.][]{mantz15}, many have relied on  
scaling relations inferred using the gas distribution alone and assuming hydrostatic equilibrium to reconstruct the cluster mass. 
Numerical simulations have shown concerns that such methods could underestimate 
cluster masses by up to 20--30\%, due to the turbulent motion and non-thermal pressure of the 
intra-cluster medium (ICM). In addition, the spread among results obtained by different groups 
indicates that the systematic uncertainties on the cluster mass calibration may currently
be underestimated \citep{rozo14,sereno15}  for X-ray and for weak lensing derived masses.

The XXL survey is an XMM Very Large Programme covering 50~$\mathrm{deg^2}$  with 
$\sim$10ks  exposures \citep[][Paper I]{xxlpaperI}. It was specifically designed to constrain 
cosmological parameters, in  particular the dark energy  (DE) equation of state through 
the combination of cluster statistics with the Planck CMB results \citep{pierre11}.
In the first series of XXL papers, our preliminary analysis, based on some 100 clusters, 
indicated that the Planck 2015 CMB cosmology overpredicts cluster counts by $\sim 20\%$ 
\citep[][hereafter Paper II]{xxlpaperII}. In the present article we perform a first complete 
cosmological analysis with a sample almost twice as large. \\
We describe the cluster sample and compare its redshift distribution with that expected 
from recent CMB measurements in the  Section 2. Section~\ref{sec:Results} presents a quantitative 
comparison between the cosmological constraints from the XXL sample and from the Planck 
CMB analysis, for a simple cosmological constant model and for a more general dark energy 
equation of state ($w=p_\mathrm{DE}/(\rho_\mathrm{DE}c^2)\ne-1$). In Section~\ref{sec:Discussion}, we discuss the 
significance of the results in view of the error budget from systematic uncertainties.  \\
For the analysis of XXL clusters in this paper, we assume a flat Universe with massless neutrinos. The number density 
of galaxy clusters follows the \citet{tinker08} mass function and the linear growth of 
cosmological overdensities is computed using version 2.6.3 of the CLASS code \citep{blas11}.

\section{ Cluster sample}
The strength of XXL resides in its well-characterised selection function, based on purely observable parameters
(X-ray flux and core radius). 
This allows us to define cluster samples with a very low contamination rate from misclassified point sources (AGN); 
see \citet{pacaud06} for a description and a graphic representation of the selection function. 
With the second release of the XXL survey, we provide a large and complete sample of 365 clusters 
\citep[][hereafter Paper~XX]{xxlpaperXX} along with various cluster measurements, including spectroscopic redshift confirmation. For statistical studies, our source selection operates in a two-dimensional parameter space combining the measured extent of the sources and the significance of this extension (the {extent statistic}, see \citealt{pacaud06}). From these data, we define a complete sub-sample of 191 sources with the highest significance of extension, located in the 47.36\,deg$^2$ of XXL data where the cluster properties can be robustly estimated, namely the C1 sample. The selection function of this sample was thoroughly estimated from Monte Carlo simulations as a function of the input flux and extent of $\beta$-model sources \citep{cavaliere76}, as explained in
\citepalias{xxlpaperII}.
In this paper, we present cosmological constraints based on 178 of these C1 clusters that have a measured redshift between 0.05 and 1.0 (all spectroscopic but one). This redshift sub-selection ensures that our analysis would not be affected by a poorly understood selection function at very low and high redshift. 
While 8 of the 13 excluded clusters indeed fall outside the redshift range, 5 actually still lack a redshift estimate.
We account for the latter in the model as a constant incompleteness factor of 6.6\% in the redshift range [0.4--1.0], thereby assuming that they would have been spectroscopically identified if their galaxies were brighter.

\begin{table*}[t]
\begin{center}
\begin{tabular}{ c c c c c c c c}
\hline
\hline
Y & X & $X_0$ & $Y_0$ & $\alpha$ & $\gamma$ & Scatter & Reference\\
\hline
M$_\mathrm{500,WL}$ & T$_{\mathrm{300kpc}}$ &  1\,keV  & (2.60$\pm$0.55)\,$\mathrm{{\times}10^{13}\,M_\odot}$  & 1.67    &  -1.0    & -  &  \citetalias{xxlpaperIV}  \\
L$^\mathrm{XXL}_\mathrm{500}$ & T$_{\mathrm{300kpc}}$ &  1\,keV  & 8.24$\mathrm{{\times}10^{41}\,erg\,s^{-1}}$  &  3.17  & 0.47$\pm$0.68  &  0.67 & \citetalias{xxlpaperXX} \\
r$_\mathrm{c}$     & r$_\mathrm{500}$    & 1\,Mpc   & 0.15\,Mpc                & 1   &  0   & -  &  \citetalias{xxlpaperII}\\
\hline
\hline
\end{tabular}
\vspace{0.1cm}
\caption{Cluster scaling relations used in the study. 
All scaling laws are modeled as a power law of the form $Y/Y_0 = (X/X_0)^\alpha E(z)^\gamma$, where E(z) is the redshift evolution of the Hubble parameter, $E(z)=H(z)/H_0$. When indicated, a log-normal scatter is included around the mean scaling relation. Errors in the $Y_0$ or $\gamma$ columns indicate the uncorrelated Gaussian priors used to fit cosmological parameters -- parameters provided without errors are held fixed. As a matter of consistency, the luminosities used for the scaling relation of \citepalias{xxlpaperXX} are extrapolated to $r_\mathrm{500}$ from measurements performed inside 300\,kpc using the same $\beta$-model as in our selection function and cosmological modelling. As appropriate the statistical model used to derive those scaling relations account for the significant Malmquist and Eddington biases affecting our sample.\vspace{-0.4cm}}
\label{scalrel}
\end{center}
\end{table*}
 
\begin{figure}
   \centering
\includegraphics[width=9cm]{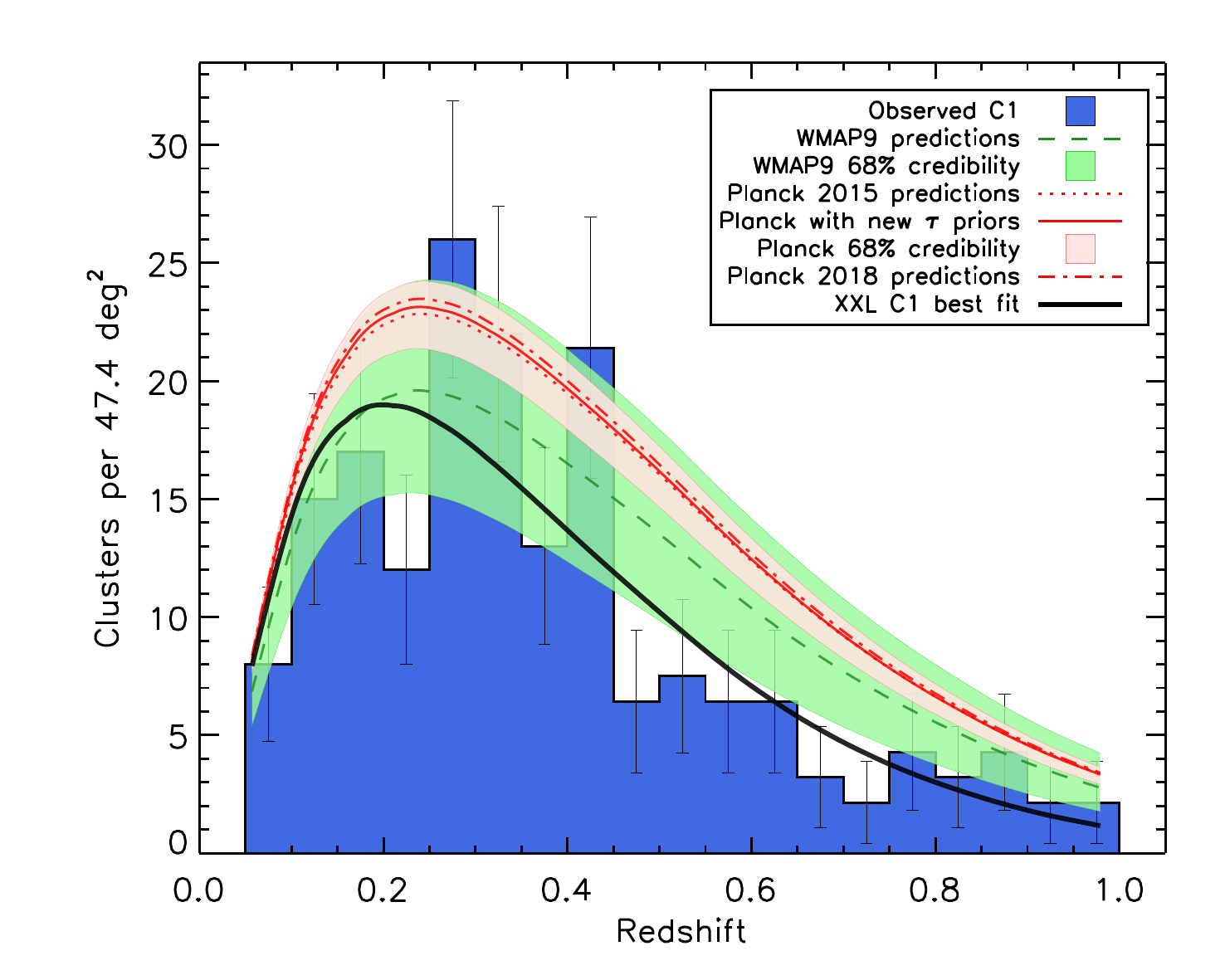}
   \caption{The histogram shows the observed redshift distribution of the 178 XXL C1 clusters used in the present study. 
   Errors bars account for shot noise and sample variance following  \citet{valageas11}; 
   the cluster deficit at $z\sim 0.5$ is present in both the XXL-N and XXL-S fields. 
   Overlaid,  the modelling obtained for different cosmologies assuming the cluster scaling 
   relations of \autoref{scalrel}. 
   The green line shows the prediction from the mean WMAP9 cosmology. 
   The red dotted line corresponds to the Planck 2015 parameters (TT+lowTEB+lensing) of \citet{planck2015XIII}. 
   The red full line shows the prediction from our reanalysis of the Planck 2015 data adopting the updated estimate of the optical depth to reionisation $\tau$ presented in \citet{planckIntermXLVI}, which we describe in \autoref{append:CMB}. For comparison we also show the prediction of the recent Planck 2018 analysis \citet{planck2018VI} which includes the final polarisation analysis (dot-dahed line).
   The shaded areas around model predictions correspond to uncertainties on the corresponding cosmological parameters, but do not include any error on scaling relations. Finally, the black thick line shows our best-fit $\Lambda$CDM model to the XXL clusters of \autoref{sec:LCDMfit}, which provide a very good fit to the data.\vspace{-0.3cm}
  }
  \label{dndz}
\end{figure}

\noindent
We show in \autoref{dndz} the redshift distribution of the C1 sample, which peaks at $z = 0.3-0.4$. 
Cluster masses are of the order of $M_\mathrm{500} \sim 10^{14}  M_{\odot}$, hence sampling a very different population than the Planck S-Z clusters \citepalias{xxlpaperII}. For comparison, we also 
display expectations from recent CMB cosmological parameter sets \citep{hinshaw13,planck2015XIII}.
These rely on three scaling relations which we use to predict cluster observational properties:
the cluster mass-to-temperature relation \citep[$M_\mathrm{500,WL}$--$T_\mathrm{300kpc}$,][hereafter Paper IV]{xxlpaperIV}, our newest 
determination of the luminosity-to-temperature relation ($L^\mathrm{XXL}_\mathrm{500}$--$T_\mathrm{300kpc}$) given in \citetalias{xxlpaperXX}, and the link between the cluster physical size $r_\mathrm{500}$ and the X-ray extent $r_\mathrm{c}$ (the core radius of a $\beta$-model with $\beta$=2/3)\footnote{See \autoref{Append:Notations} for a description of the notations used for different cluster quantities in the XXL survey.}. 
The coefficients of the scaling relations used in this paper are summarised in \autoref{scalrel}.
We note that this mass calibration relies entirely on weak lensing measurements \citepalias{xxlpaperIV}.
More details on the computation of the expected cluster counts are provided in \autoref{append:Like}.

The mismatch between the XXL cluster number counts and the Planck CMB cosmology suggested by our preliminary analysis in \citetalias{xxlpaperII} remains. The predictions from WMAP9 constitute a better fit, but in both cases a slight deficit of C1 clusters is observed in the redshift range [0.4--0.7], as already reported from the analysis of a 11\,deg$^2$ subfield by \citet{clerc14}. This global deficit is also the reason for the apparent negative evolution of the cluster luminosity function discussed in \citetalias{xxlpaperXX}. Due to the better match with WMAP9, we infer that, as for the Planck sample of Sunyaev-Zel'dovich clusters, the XXL C1 sample probably favours a lower value of $\sigma_8$ than the Planck CMB cosmology. We quantitatively analyse this hypothesis in the next section.

\section{Detailed cosmological modelling}
\label{sec:Results}
\subsection{Assumptions and methods}
\label{sec:Methods}
We have run a stand-alone cosmological fit of the XXL C1 redshift distribution based on a standard Markov chain Monte Carlo procedure 
(the Metropolis algorithm). For the whole analysis, we only rely on the cluster redshifts and never 
use directly the additional information contained in the mass distribution of galaxy clusters; clusters masses only appear 
in the selection function as encoded in the scaling relations \citepalias{xxlpaperII}. Our model uses at most six free cosmological parameters: $h$, $\Omega_m$, $\Omega_b$, $\sigma_8$, $n_s$, and $w$. In most cases the dark energy equation of state parameter $w$ is  fixed to $-1$ (flat $\Lambda$CDM). 
Also included as nuisance parameters are  the optical depth to reionisation ($\tau$) in the CMB analysis,  the normalisation of the $M_\mathrm{500,WL}$--$T_\mathrm{300kpc}$ scaling relation, and the evolution of the $L^\mathrm{XXL}_\mathrm{500}$--$T_\mathrm{300kpc}$ for the XXL clusters (see \autoref{scalrel}); these parameters are then marginalised over.
Since XXL clusters are not enough  by themselves to  constrain all cosmological parameters 
(in particular, $\Omega_b$ and $n_s$ to which the cluster number density is not very sensitive),
we apply Gaussian priors on the C1-only constraints, derived from the Planck 2015 measurements (so that
the priors do not introduce any artificial mismatch between XXL and Planck) and with errors 
increased by a factor of 5 (so the priors are loose enough to not force agreement). 
We apply this to the  parameter combinations that naturally describe the BAO peak 
pattern observed in the CMB data, namely: $n_s = 0.965 \pm 0.023$, $\Omega_b h^2 = 0.0222 \pm 0.0011$, and 
$\Omega_m h^2= 0.1423 \pm 0.0073$.
In addition, we impose a conservative Gaussian prior on the Hubble constant to match observations of the 
local Universe as $h=0.7\pm0.1$.

\subsection{ \texorpdfstring{$\Lambda$}{L}CDM }
\label{sec:LCDMfit}
\begin{figure*}
 \begin{center}
 \includegraphics[width=6cm]{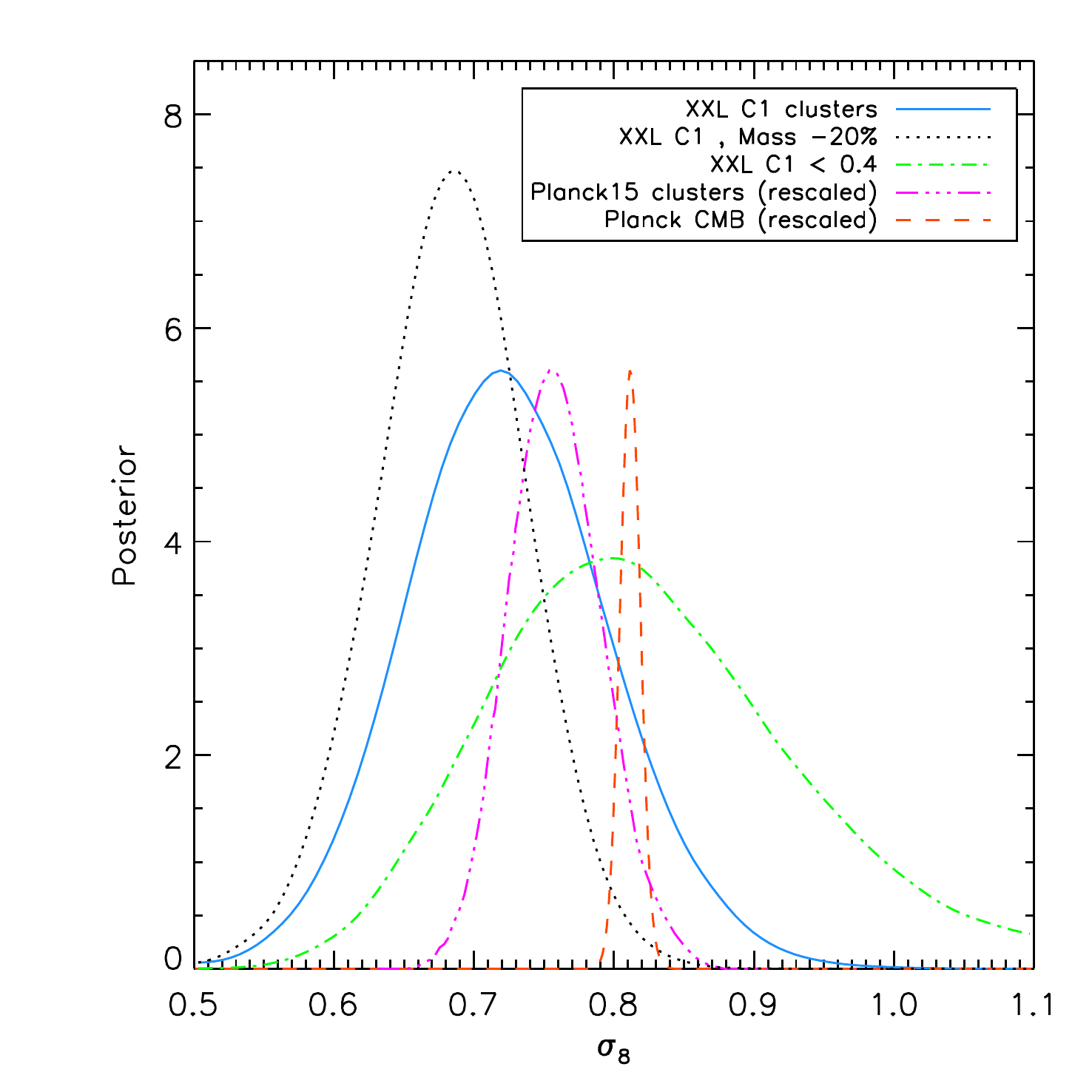}
 \includegraphics[width=6cm]{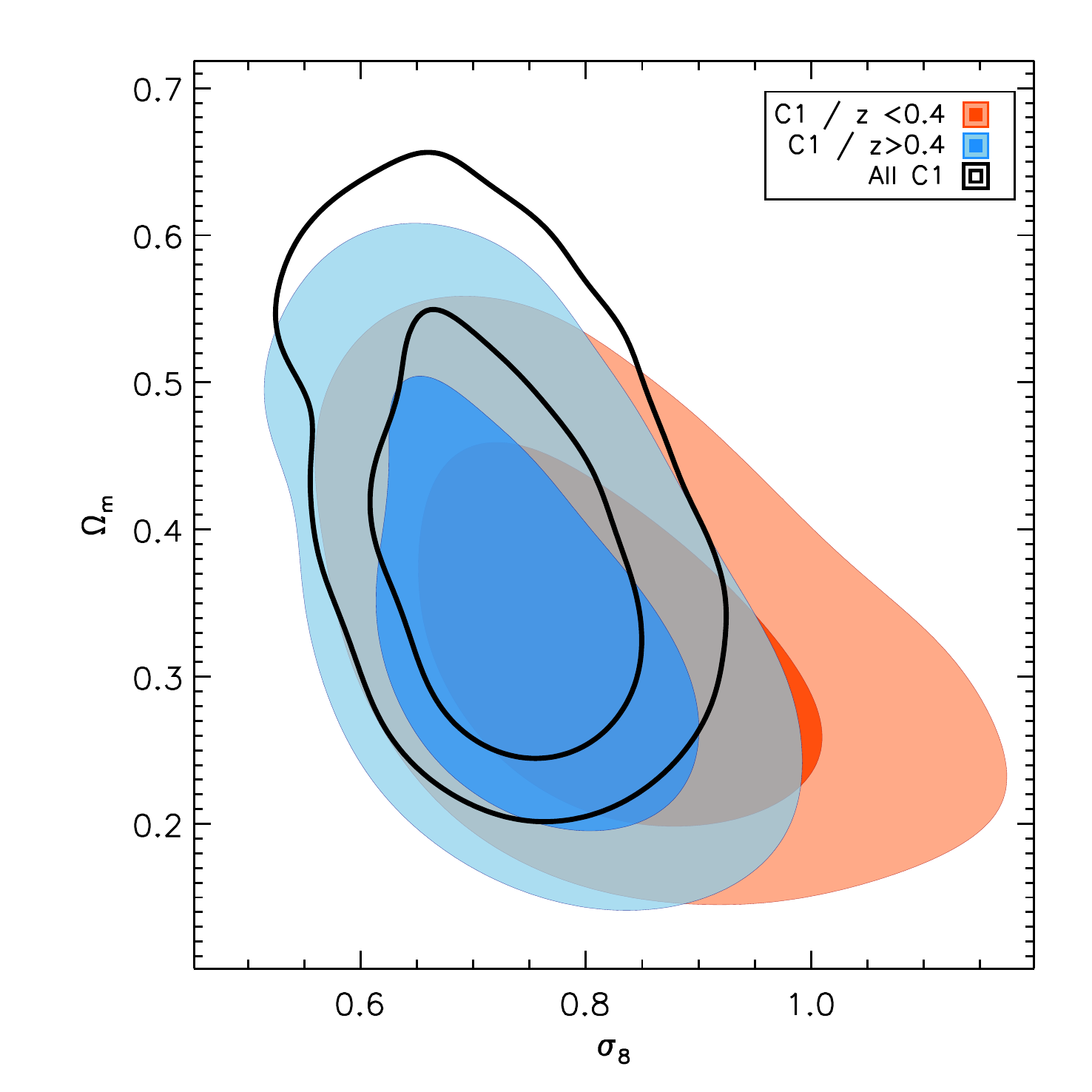}
 \includegraphics[width=6cm]{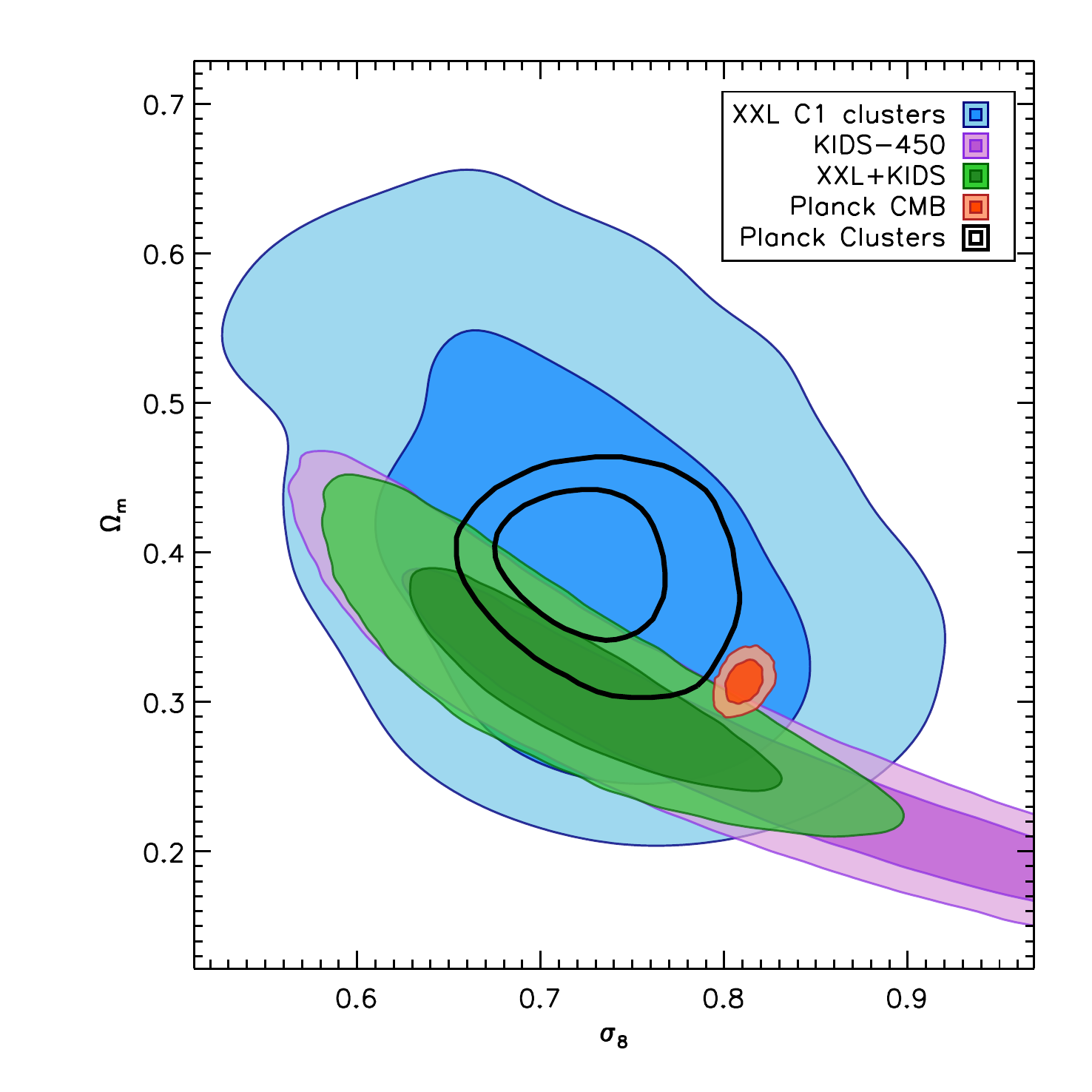}
   \caption{Cosmological constraints in the flat $\Lambda$CDM model. {Left}: Posterior distribution on 
   $\sigma_8$ from the cosmological fit of the whole XXL C1 cluster sample (blue line), when lowering the mass calibration by 20\% (black dotted line), when using only clusters below $z=0.4$ (green dot-dashed line),  for Planck clusters \citep[][pink triple dot-dashed line]{planck2015XXIV}, and for 
   CMB \citep[][orange dashed line]{planck2015XIII}, rescaled to match the peak of the XXL C1 distribution. 
   {Middle}: Countours of 1$\sigma$ and 2$\sigma$  in the $\sigma_8$--$\Omega_m$ plane obtained from the C1 clusters as a 
   function of redshift. {Right}: Comparison of the XXL, KIDS (lensing), and Planck 2015 constraints in 
   the $\sigma_8$--$\Omega_m$ plane (1$\sigma$ and 2$\sigma$ contours).}
 \label{sigma8}
 \end{center}
\end{figure*}

The results for a fixed $w=-1$ are shown in \autoref{sigma8} and compared with the constraints 
from Planck 2015 and a weak lensing tomography analysis from the KIDS survey \citep{hildebrandt17}. 
A good overlap is found between the XXL and Planck constraints; using the Index of Inconsistency (IOI, see \autoref{append:IOI}) we can quantify the significance of the offset between the two posteriors to be lower than 0.05$\sigma$. Although statistically consistent, the XXL constraints indicate a lower value of $\sigma_8$ of $0.72\pm0.07$ (versus $0.811\pm0.007$ for Planck) and a correspondingly higher value of $\Omega_m=0.40\pm0.09$ (versus $0.313\pm0.009$ for Planck). While the combination with KIDS points to a better agreement with Planck on the matter density, $\sigma_8$ remains much lower ($0.72\pm0.06$). For the XXL constraints to exactly match the Planck predictions, we would need to assume that our current masses estimates are biased by $18\pm5$ toward lower masses. This is still allowed by the current uncertainty of our $M_\mathrm{500}$--$T_\mathrm{300kpc}$ calibration, and it explains the lack of significant tension between the two datasets.  This estimate of the bias stems from the marginalised constraints on the normalisation of the $M_\mathrm{500}$--$T_\mathrm{300kpc}$ relation obtained through the combination of XXL and Planck leaving the normalisation of the relation entirely free.   

Given the apparent lack of intermediate redshift C1 clusters compared to cosmological predictions 
(\autoref{dndz}), we also investigated separately the constraints arising from C1 clusters
below and above z=0.4. As can be seen in \autoref{sigma8}, low-redshift C1 clusters show numbers consistent with the Planck CMB cosmology (although with large errors), while high-z clusters require lower values of $\sigma_8$. Since a high matter density is required to reproduce the strong redshift evolution of the full sample, the $\Omega_m$--$\sigma_8$ degeneracy conspires to push $\sigma_8$ even lower when the two redshift ranges are combined.

In a flat universe with a cosmological constant, the CMB acoustic scale sets tight  constraints on the Hubble constant, while the CMB peaks mostly fix the baryon ($\Omega_b \times h^2$), matter ($\Omega_m \times h^2$), and photon densities (through CMB black-body temperature, $T_{CMB}$): there is no strong degeneracy between the parameters.
However, when letting $\Omega_k$ or $w$ be free, the geometrical degeneracy sets in and the Planck constraints loosen drastically, leaving room for the XXL clusters to improve on the Planck CMB constraints. We investigate this possibility below.

\subsection{Dark energy}

\begin{figure*}
\begin{center}
\includegraphics[width=6cm]{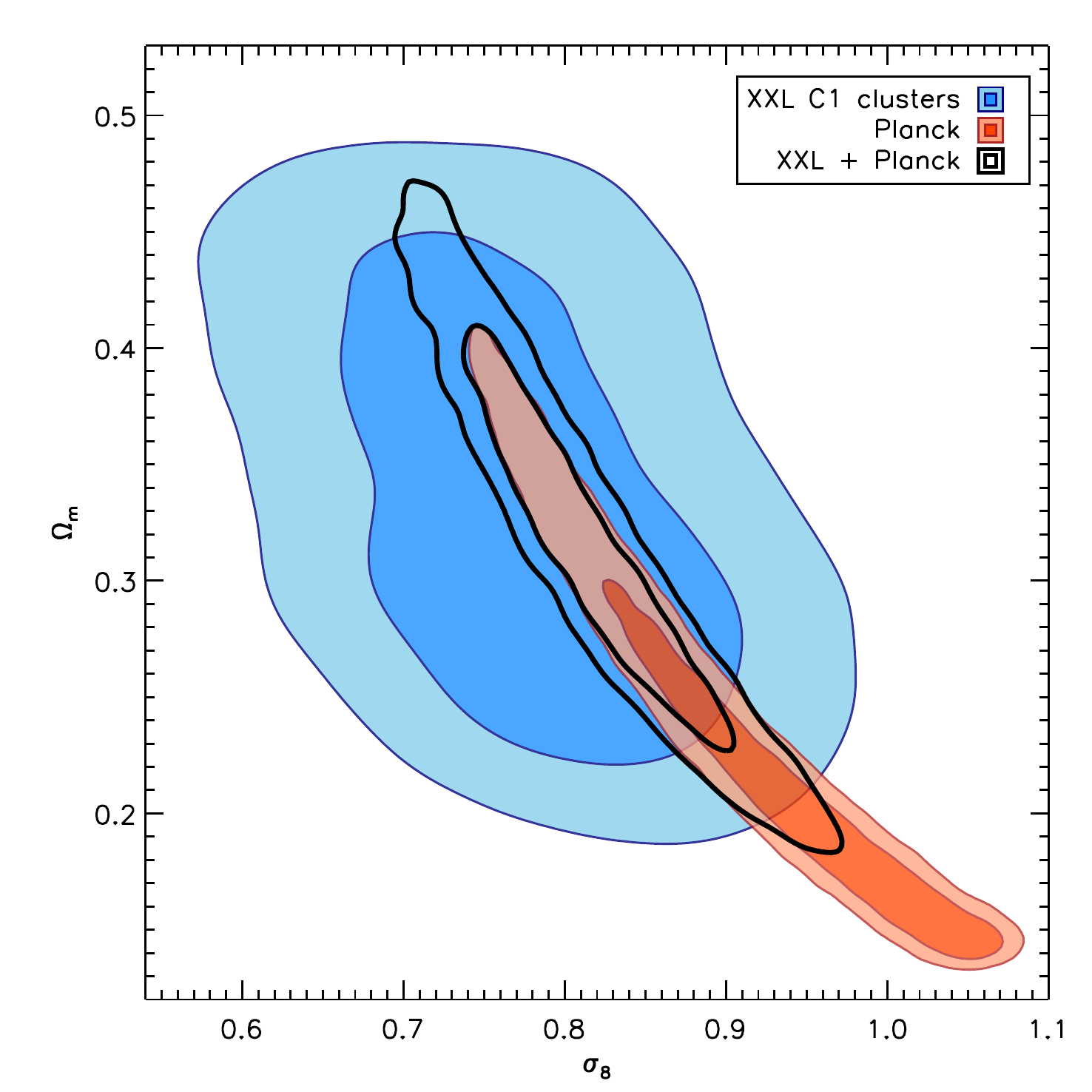}
\includegraphics[width=6cm]{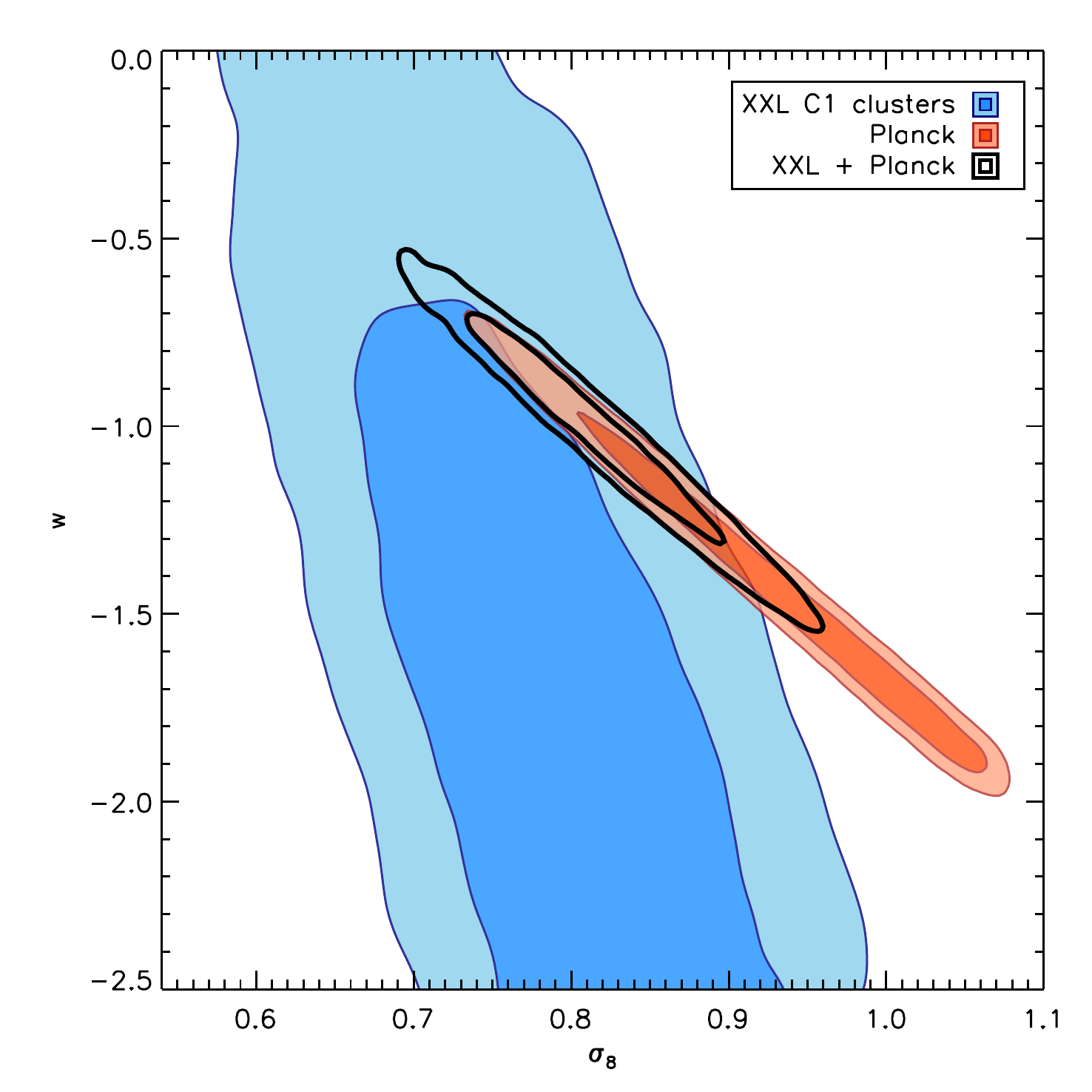} 
\includegraphics[width=6cm]{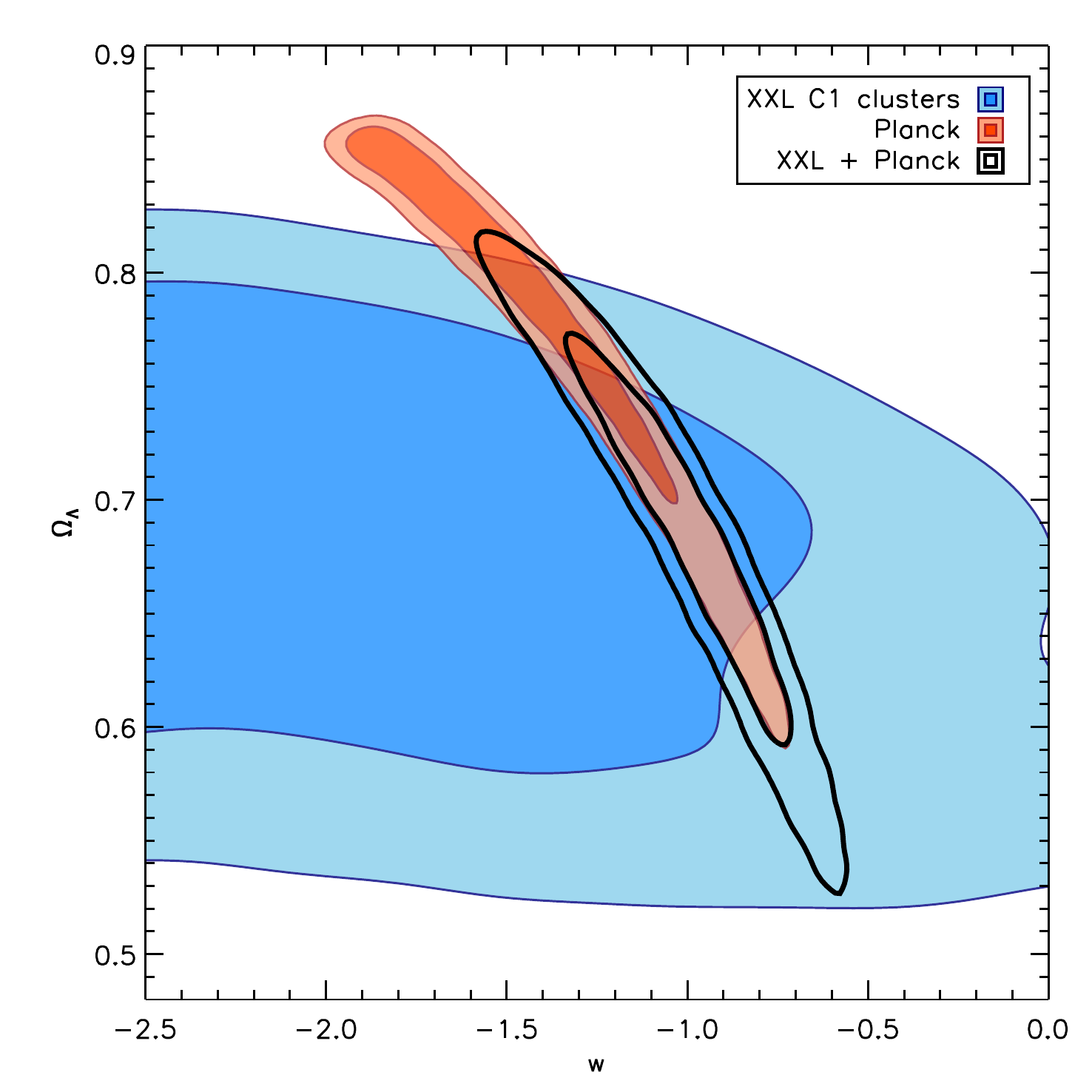} 
\caption{Comparison between the XXL and Planck 2015 constraints, with $w$ free, for the
$\sigma_{8}$--$\Omega_{m}$, $\sigma_{8}$--$w$, and $w$--$\Omega_\Lambda$ planes (1$\sigma$ and 2$\sigma$ contours,
same assumption on $\tau$ as in \autoref{dndz}).  
\vspace{-0.3cm}}
\label{cosmo}
\end{center}
\end{figure*}

The effect of releasing the value of $w$ on the Planck CMB is shown in \autoref{cosmo} for $\sigma_8$ and $\Omega_m$: 
the size of the error bars now approaches that from the XXL cluster sample, which are only slightly larger than for fixed $w$. 
The XXL dataset, like Planck, favours a strongly negative equation of state parameter (respectively  $w_0=-1.53\pm0.62$ and  $-1.44\pm0.30$) albeit with rather different values for the other parameters. 
Actually, most of the larger parameter space now allowed by the CMB datasets is disfavoured by the XXL C1 clusters, which thus hold the potential to improve significantly on  the dark energy constraints provided by Planck alone. 
Still, the constraints obtained from both projects show good overlap, and our inconsistency test with the IOI shows that the two datasets are compatible within $\sim0.5\sigma$ (PTE=0.49). 
In the absence of any apparent tension between the two probes, we thus proceed with their combination.

The joint C1+Planck dataset results in a significantly higher value of $w=-1.02\pm0.20$, than would each probe if  taken separately, with a best-fit cosmology similar to the preferred Planck ${\Lambda}$CDM model. Interestingly, other datasets, like supernovae \citep[][Fig. 14]{betoule14}, also favour equation of state parameters that differ from $-1$ but, once combined with Planck or other probes, point toward the concordance $\Lambda$CDM cosmology. In addition to comforting the $\Lambda$CDM model, our cosmological analysis of the XXL C1 cluster decreases by 30\% the errors on $w$ obtained from Planck alone, despite using less than half of the final cluster sample and neglecting the constraints provided by both the mass distribution and spatial correlation of clusters.

\section{Discussion and conclusion}
\label{sec:Discussion}
 
All in all, our results prove consistent with the Planck S-Z cluster analysis \citep{planck2015XXIV} despite relying on a totally different cluster dataset (mass and redshift range, selection procedure, scaling relations based on weak lensing mass measurements). However, the uncertainties resulting from the present analysis are too large to either confirm or dismiss the tension identified within the Planck collaboration between the primary CMB and the abundance of galaxy clusters. 
Since our analysis relied on less than half of the full XXL cluster sample, did not use information from the cluster mass distribution and assumed conservative errors on scaling relations, there is ample room for improvement in the constraints provided by XXL alone in the near future.
Yet, we showed that, even at the present stage, the XXL clusters already bring significant improvements on dark energy constraints when combined with Planck data. 

While most of the critics of the Planck sample analysis pertains to the hydrostatic bias and its normalisation via numerical simulations, this does not directly affect the present studies which relies on weak lensing mass measurements. There are nevertheless a number of residual uncertainties in the present analysis, in particular regarding our mass estimates, which need to be addressed before using the full power of the survey, and that we now discuss. 
Key considerations for the XXL analysis include the following:
\begin{itemize}
\item[•] \underline{Accuracy of the mass calibration}: in \citet[XXL Paper XIII]{xxlpaperXIII}, 
         we analysed the gas mass of 100 XXL galaxy clusters and found that their gas mass fractions
         were about 20\% lower than expected. A possible interpretation would be for the mass 
         calibration published in \citetalias{xxlpaperIV} to be ovestimated by $ \sim$20\%, which 
         is supported by the parallel weak lensing analysis presented in \citet{lieu17}. 
         To test the impact of such a calibration offset, we repeated our flat $\Lambda$CDM analysis
         decreasing the prior on the $M_\mathrm{500,WL}$--$T_\mathrm{300kpc}$ normalisation by 20\%. 
         In this case, the XXL clusters would start to deviate more significantly from the prediction of the Planck CMB (by $\sim 1.1\sigma$) with marginalised values of $\sigma_8=0.68\pm0.05$ and $\Omega_m=0.35\pm0.08$. 
         The marginalised posterior on $\sigma_8$ for this case is also shown in the left panel of \autoref{sigma8}.
\item[•] \underline{Scaling relation model}: for our analysis, we have assumed 
                 a bijective relation between cluster mass and temperature, and have attributed all the scatter 
                 in cluster scaling laws to the relation between temperature and luminosity. A more 
                 realistic model is required that would include the scatter in both luminosity and 
                 temperature, as well as their covariance. Another option would be to bypass 
                 the need for cluster temperatures by estimating the luminosity in a redshifted band corresponding to the measured flux. Only one scaling relation and its redshift evolution would then be required without covariance issue.
         In addition, an accurate cosmological analysis requires reevaluating simultaneously cluster 
         scaling relations as the cosmology is varied \citep[e.g.][]{mantz10a}. Given the 
         large uncertainties in the present analysis, this was not considered necessary,
         but the same will no longer hold for studies with more clusters and better
         mass information. 
\item[•] \underline{The average shape of galaxy clusters}: in the scaling model of \autoref{scalrel},
                 we chose a specific model for the surface brightness of galaxy clusters (a $\beta$ = 2/3 model and 
                 $x_\mathrm{c}=r_\mathrm{c}/r_\mathrm{500}=0.15$). Although motivated by observations, the value of 
                 $x_\mathrm{c}$ is not firmly established, in particular in the new mass--redshift regime
                 uncovered by XXL. Most other plausible values of $x_\mathrm{c}$ would lower the number of expected clusters
                 and improve the agreement with the Planck CMB model: the detection efficiency becomes lower 
                 for  very compact clusters (which may be misclassified as X-ray active galactic nuclei, AGN) and
                 for very extended clusters (whose low surface brightness hampers their detection).
                 As for the normalisation of $M_\mathrm{500,WL}$--$T_\mathrm{300kpc}$ in \autoref{sec:Results}, we estimated the value of
                 $x_\mathrm{c}$ required by the Planck CMB data from its marginalised constraints when
                 Planck and XXL are combined. The resulting constraints on $x_\mathrm{c}$ are surprisingly 
                 loose, indicating that $x_\mathrm{c}$ is not a major systematic in the present study. Furthermore,
                 our fiducial value of 0.15 is only 1.1$\sigma$ away from the preferred value of 
                 $x_\mathrm{c}=0.44\pm0.26$ so that changing this parameter cannot improve  the agreement 
                 between XXL and Planck much.
\item[•] \underline{The dispersion of cluster shapes}: The fourth paper in the ASpiX series 
         \citep{valotti18} studies the impact of introducing some scatter around $x_\mathrm{c}$ in 
         the relation between $r_\mathrm{c}$ and $r_\mathrm{500}$. 
         Although the results depend on the exact value of $x_\mathrm{c}$, a larger scatter usually implies 
         fewer detected clusters. Using again the same method, we found that the combination of
         XXL C1 + Planck implies a log-normal scatter of $1.49\pm0.31$. This value is 
         well constrained, showing that an increase in the scatter could, in principle, 
         change the interpretation of the results. We will pay greater attention to this parameter
         in the forthcoming analyses;  in the meantime, we note that the preferred value above
         is unlikely to be realistic as the gas distribution in galaxy clusters is observed 
         to be rather self-similar \citep[e.g.][]{croston08} and numerical simulations predict
         a much lower scatter (for instance \citealt{lebrun17} and \citealt{valotti18} estimated 
          a log-normal scatter of 0.5 on $x_\mathrm{c}$ from the OWLS simulations).
\item[•] \underline{The effect of peaked clusters}: as noted by \citet{clerc14}, a change with
         redshift in the strength or frequency of cool cores, as well as a different occupation
         of cluster halos by AGNs, could explain the apparent deficit of clusters
         at intermediate redshift. So far, our observational programme to identify clusters 
         contaminated by AGNs proved that the C1 selection is robust 
         \citep[][XXL paper XXXIII]{xxlpaperXXXIII}. 
         However, we already noticed that AGNs may be more common in the centre  of the
         XXL groups than they are in low-redshift massive clusters 
         \citep[][XXL paper XXXV]{xxlpaperXXXV}. 
         In the future, we will use realistic simulations of the combined cluster and AGN populations obtained in \citet[][XXL paper XIX]{xxlpaperXIX} to further 
         investigate these hypotheses.
\item[•] \underline{Systematics of theoretical mass functions}:  here we rely on the commonly 
         used \citet{tinker08} mass function, but over the years a number of new results have become available \citep[e.g.][]{watson13,despali16} that use higher resolution 
         simulations and better statistics. Differences still remain  between them, which means that 
         an estimate of systematic uncertainties impinging on the mass function 
         itself must be included. Even more importantly, results from magneto-hydrodynamic
         simulations have shown that the detailed physics of the gas affects the collapse of 
         dark matter halos and alters the mass function \citep{stanek09}. 
         Even though analytical recipes already exist to include this effect 
         \citep[e.g.][]{velliscig14,bocquet16}, there is still enough uncertainty in the 
         modelling of the gas that results still vary between different simulations and codes.\vspace{-0.2cm}
\end{itemize}  

To conclude, the present article constitutes a significant step in the cosmological analysis of X-ray cluster samples, targeting a mass and redshift range that will be the realm of wide-area upcoming surveys (ACT-pol, SPT-pol, eRosita, Euclid).
When the final XXL data release occurs, a more comprehensive study  involving the full cluster sample (some 400 objects) will follow, and will address  most of the shortcomings noted above. 
Our cosmology pipeline will be upgraded to jointly fit cosmology and scaling relations relying directly on the observed signal. One such observable will be the angular extent of clusters for which a scaling relation and scatter will be constrained simultaneously. In parallel, the selection function will undergo significant tests based on realistic MHD simulations \citepalias{xxlpaperI,xxlpaperXIX} to assess the effect of cool cores and AGN contamination. In addition, lowering our threshold on the extent statistic (i.e. using the  C2 sample described in \citealt{xxlpaperXX}) will roughly double the number of clusters and should improve the cosmological constraints by a factor of $\sim\sqrt{2}$ \citep{pierre11};  the new clusters correspond to lower S/N sources, hence to less massive  or more distant clusters.
The calibration of the scaling relations will also improve, thanks to lensing mass measurements by the HSC at the Subaru telescope. We shall thus  be in a position to model the $dn/dM/dz$ distribution (much more constraining than $dn/dz$) in combination with the final Planck chains \citep{planck2018VI}.
The final results will be combined with those from the 3D XXL cluster-cluster correlation function obtained with the same sample \citep[][XXL Paper XVI]{xxlpaperXVI}; when $w$ is free, this combination has the potential to double the precision on the DE equation of state \citep{pierre11}.
All in all, by combining a better mass determination, the information from the mass function, the increase in sample size, and the correlation function we expect an improvement of a factor of 3 with respect to the current analysis. We will also be in a strong position to quantify the agreement between XXL and the Planck CMB results: dividing by 3 the current XXL cosmological constraints while keeping the same best-fit model would  result for instance in a 4.8$\sigma$ and 13.4$\sigma$ tension, respectively in the $\mathrm{\Lambda}$CDM and wCDM models based on our IOI test.


\begin{acknowledgements}
XXL is an international project based on an XMM Very Large Programme
surveying two 25 \dd\ extragalactic fields at a depth of
$\sim6\times10^{-15}\,$\flux\ in $[$0.5--2$]$ keV.
The XXL website is {\tt http://irfu.cea.fr/xxl}.
Multiband information and spectroscopic follow-up of the X-ray sources are
obtained through a number of survey programmes, summarised at
{\tt http://xxlmultiwave.pbworks.com/}. FP acknowledges support by the German Aerospace Agency (DLR) with funds from the Ministry of Economy and Technology (BMWi) through grant 50 OR 1514, and thanks Thomas Reiprich for providing this financial support. 
EK acknowledges the Centre National dEtudes Spatiales (CNES) and CNRS for supporting his post-doctoral research.
The Saclay group also acknowledges long-term support from the CNES. 
Finally, we want to thank Dominique Eckert, Stefano Ettori, Nicolas Clerc, Pier Stefano Corasaniti, Hendrik Hildebrandt, Amandine Le Brun, Ian McCarthy, David Rapetti, and Marina Ricci for useful comments and discussions on the content of this article.
\end{acknowledgements}

\bibliographystyle{aa} 
\bibliography{paperlib}{}

\begin{thebibliography}{50}
\expandafter\ifx\csname natexlab\endcsname\relax\def\natexlab#1{#1}\fi

\bibitem[{{Adami} {et~al.}(2018){Adami}, {Giles}, {Koulouridis}, {Pacaud},
  {Caretta}, {Pierre}, {Eckert}, {Ramos-Ceja}, {Gastaldello}, {Fotopoulou},
  {Guglielmo}, {Lidman}, {Sadibekova}, {Iovino}, {Maughan}, {Chiappetti},
  {Alis}, {Altieri}, {Baldry}, {Bottini}, {Birkinshaw}, {Bremer}, {Brown},
  {Cucciati}, {Driver}, {Elmer}, {Ettori}, {Evrard}, {Faccioli}, {Granett},
  {Grootes}, {Guzzo}, {Hopkins}, {Horellou}, {Lef\`evre}, {Liske}, {Malek},
  {Marulli}, {Maurogordato}, {Owers}, {Paltani}, {Poggianti}, {Polletta},
  {Plionis}, {Pollo}, {Pompei}, {Ponman}, {Rapetti}, {Ricci}, {Robotham},
  {Tuffs}, {Tasca}, {Valtchanov}, {Vergani}, {Wagner}, {Willis}, \& {XXL
  collaboration}}]{xxlpaperXX}
{Adami}, C., {Giles}, P., {Koulouridis}, E., {et~al.} 2018, \aap, in press, DOI
  10.1051/0004-6361/201731606 (XXL Paper XX)

\bibitem[{{Betoule} {et~al.}(2014){Betoule}, {Kessler}, {Guy}, {Mosher},
  {Hardin}, {Biswas}, {Astier}, {El-Hage}, {Konig}, {Kuhlmann}, {Marriner},
  {Pain}, {Regnault}, {Balland}, {Bassett}, {Brown}, {Campbell}, {Carlberg},
  {Cellier-Holzem}, {Cinabro}, {Conley}, {D'Andrea}, {DePoy}, {Doi}, {Ellis},
  {Fabbro}, {Filippenko}, {Foley}, {Frieman}, {Fouchez}, {Galbany}, {Goobar},
  {Gupta}, {Hill}, {Hlozek}, {Hogan}, {Hook}, {Howell}, {Jha}, {Le Guillou},
  {Leloudas}, {Lidman}, {Marshall}, {M{\"o}ller}, {Mour{\~a}o}, {Neveu},
  {Nichol}, {Olmstead}, {Palanque-Delabrouille}, {Perlmutter}, {Prieto},
  {Pritchet}, {Richmond}, {Riess}, {Ruhlmann-Kleider}, {Sako}, {Schahmaneche},
  {Schneider}, {Smith}, {Sollerman}, {Sullivan}, {Walton}, \&
  {Wheeler}}]{betoule14}
{Betoule}, M., {Kessler}, R., {Guy}, J., {et~al.} 2014, \aap, 568, A22

\bibitem[{{Blas} {et~al.}(2011){Blas}, {Lesgourgues}, \& {Tram}}]{blas11}
{Blas}, D., {Lesgourgues}, J., \& {Tram}, T. 2011, \jcap, 7, 034

\bibitem[{{Bocquet} {et~al.}(2016){Bocquet}, {Saro}, {Dolag}, \&
  {Mohr}}]{bocquet16}
{Bocquet}, S., {Saro}, A., {Dolag}, K., \& {Mohr}, J.~J. 2016, \mnras, 456,
  2361

\bibitem[{{Cavaliere} \& {Fusco-Femiano}(1976)}]{cavaliere76}
{Cavaliere}, A. \& {Fusco-Femiano}, R. 1976, \aap, 49, 137

\bibitem[{{Clerc} {et~al.}(2014){Clerc}, {Adami}, {Lieu}, {Maughan}, {Pacaud},
  {Pierre}, {Sadibekova}, {Smith}, {Valageas}, {Altieri}, {Benoist},
  {Maurogordato}, \& {Willis}}]{clerc14}
{Clerc}, N., {Adami}, C., {Lieu}, M., {et~al.} 2014, \mnras, 444, 2723

\bibitem[{{Croston} {et~al.}(2008){Croston}, {Pratt}, {B{\"o}hringer},
  {Arnaud}, {Pointecouteau}, {Ponman}, {Sanderson}, {Temple}, {Bower}, \&
  {Donahue}}]{croston08}
{Croston}, J.~H., {Pratt}, G.~W., {B{\"o}hringer}, H., {et~al.} 2008, \aap,
  487, 431

\bibitem[{{Despali} {et~al.}(2016){Despali}, {Giocoli}, {Angulo}, {Tormen},
  {Sheth}, {Baso}, \& {Moscardini}}]{despali16}
{Despali}, G., {Giocoli}, C., {Angulo}, R.~E., {et~al.} 2016, \mnras, 456, 2486

\bibitem[{{Eckert} {et~al.}(2016){Eckert}, {Ettori}, {Coupon}, {Gastaldello},
  {Pierre}, {Melin}, {Le Brun}, {McCarthy}, {Adami}, {Chiappetti}, {Faccioli},
  {Giles}, {Lavoie}, {Lef{\`e}vre}, {Lieu}, {Mantz}, {Maughan}, {McGee},
  {Pacaud}, {Paltani}, {Sadibekova}, {Smith}, \& {Ziparo}}]{xxlpaperXIII}
{Eckert}, D., {Ettori}, S., {Coupon}, J., {et~al.} 2016, \aap, 592, A12, (XXL
  Paper XIII)

\bibitem[{{Gelman} \& {Rubin}(1992)}]{gelman92}
{Gelman}, A. \& {Rubin}, D.~B. 1992, Statistical Science, 7, 457

\bibitem[{{Hildebrandt} {et~al.}(2017){Hildebrandt}, {Viola}, {Heymans},
  {Joudaki}, {Kuijken}, {Blake}, {Erben}, {Joachimi}, {Klaes}, {Miller},
  {Morrison}, {Nakajima}, {Verdoes Kleijn}, {Amon}, {Choi}, {Covone}, {de
  Jong}, {Dvornik}, {Fenech Conti}, {Grado}, {Harnois-D{\'e}raps}, {Herbonnet},
  {Hoekstra}, {K{\"o}hlinger}, {McFarland}, {Mead}, {Merten}, {Napolitano},
  {Peacock}, {Radovich}, {Schneider}, {Simon}, {Valentijn}, {van den Busch},
  {van Uitert}, \& {Van Waerbeke}}]{hildebrandt17}
{Hildebrandt}, H., {Viola}, M., {Heymans}, C., {et~al.} 2017, \mnras, 465, 1454

\bibitem[{{Hinshaw} {et~al.}(2013){Hinshaw}, {Larson}, {Komatsu}, {Spergel},
  {Bennett}, {Dunkley}, {Nolta}, {Halpern}, {Hill}, {Odegard}, {Page}, {Smith},
  {Weiland}, {Gold}, {Jarosik}, {Kogut}, {Limon}, {Meyer}, {Tucker}, {Wollack},
  \& {Wright}}]{hinshaw13}
{Hinshaw}, G., {Larson}, D., {Komatsu}, E., {et~al.} 2013, \apjs, 208, 19

\bibitem[{{Israel} {et~al.}(2015){Israel}, {Schellenberger}, {Nevalainen},
  {Massey}, \& {Reiprich}}]{israel15}
{Israel}, H., {Schellenberger}, G., {Nevalainen}, J., {Massey}, R., \&
  {Reiprich}, T.~H. 2015, \mnras, 448, 814

\bibitem[{Jeffreys(1961)}]{jeffreys61}
Jeffreys, H. 1961, The theory of probability (Oxford Universiy Press)

\bibitem[{{Joudaki} {et~al.}(2017){Joudaki}, {Blake}, {Heymans}, {Choi},
  {Harnois-Deraps}, {Hildebrandt}, {Joachimi}, {Johnson}, {Mead}, {Parkinson},
  {Viola}, \& {van Waerbeke}}]{joudaki17}
{Joudaki}, S., {Blake}, C., {Heymans}, C., {et~al.} 2017, \mnras, 465, 2033

\bibitem[{{Koulouridis} {et~al.}(2018{\natexlab{a}}){Koulouridis}, {Faccioli},
  {Le Brun}, {Plionis}, {McCarthy}, {Pierre}, {Akylas}, {Georgantopoulos},
  {Paltani}, {Lidman}, {Fotopoulou}, {Vignali}, {Pacaud}, \&
  {Ranalli}}]{xxlpaperXIX}
{Koulouridis}, E., {Faccioli}, L., {Le Brun}, A.~M.~C., {et~al.}
  2018{\natexlab{a}}, \aap, in press, DOI 10.1051/0004-6361/201730789 (XXL
  paper XIX), arXiv eprints [\eprint[arXiv]{1709.01926}]

\bibitem[{{Koulouridis} {et~al.}(2018{\natexlab{b}}){Koulouridis}, {Ricci},
  {Giles}, {Adami}, {Faccioli}, {Ramos-Ceja}, {Pierre}, {Plionis}, {Lidman},
  {Georgantopoulos}, {Chiappetti}, {Elyiv}, {Ettori}, {Fotopoulou},
  {Gastaldello}, {Pacaud}, {Paltani}, \& {Vignali}}]{xxlpaperXXXV}
{Koulouridis}, E., {Ricci}, M., {Giles}, P., {et~al.} 2018{\natexlab{b}}, \aap,
  in press, DOI 10.1051/0004-6361/201832974 (XXL paper XXXV), arXiv eprints
  [\eprint[arXiv]{1809.00683}]

\bibitem[{{Le Brun} {et~al.}(2017){Le Brun}, {McCarthy}, {Schaye}, \&
  {Ponman}}]{lebrun17}
{Le Brun}, A.~M.~C., {McCarthy}, I.~G., {Schaye}, J., \& {Ponman}, T.~J. 2017,
  \mnras, 466, 4442

\bibitem[{{Lieu} {et~al.}(2017){Lieu}, {Farr}, {Betancourt}, {Smith}, {Sereno},
  \& {McCarthy}}]{lieu17}
{Lieu}, M., {Farr}, W.~M., {Betancourt}, M., {et~al.} 2017, \mnras, 468, 4872

\bibitem[{{Lieu} {et~al.}(2016){Lieu}, {Smith}, {Giles}, {Ziparo}, {Maughan},
  {D{\'e}mocl{\`e}s}, {Pacaud}, {Pierre}, {Adami}, {Bah{\'e}}, {Clerc},
  {Chiappetti}, {Eckert}, {Ettori}, {Lavoie}, {Le Fevre}, {McCarthy},
  {Kilbinger}, {Ponman}, {Sadibekova}, \& {Willis}}]{xxlpaperIV}
{Lieu}, M., {Smith}, G.~P., {Giles}, P.~A., {et~al.} 2016, \aap, 592, A4, (XXL
  Paper IV)

\bibitem[{{Lin} \& {Ishak}(2017)}]{lin17}
{Lin}, W. \& {Ishak}, M. 2017, \prd, 96, 023532

\bibitem[{{Logan} {et~al.}(2018){Logan}, {Maughan}, {Bremer}, {Giles},
  {Birkinshaw}, {Chiappetti}, {Clerc}, {Faccioli}, {Koulouridis}, {Pacaud},
  {Pierre}, {Ramos-Ceja}, \& {Willis}}]{xxlpaperXXXIII}
{Logan}, C.~H.~A., {Maughan}, B.~J., {Bremer}, M., {et~al.} 2018, \aap, in
  press, DOI 10.1051/0004-6361/201833654 (XXL paper XXXIII), arXiv eprints
  [\eprint[arXiv]{1808.04623}]

\bibitem[{{Mantz} {et~al.}(2010){Mantz}, {Allen}, {Rapetti}, \&
  {Ebeling}}]{mantz10a}
{Mantz}, A., {Allen}, S.~W., {Rapetti}, D., \& {Ebeling}, H. 2010, \mnras, 406,
  1759

\bibitem[{{Mantz} {et~al.}(2015){Mantz}, {von der Linden}, {Allen},
  {Applegate}, {Kelly}, {Morris}, {Rapetti}, {Schmidt}, {Adhikari}, {Allen},
  {Burchat}, {Burke}, {Cataneo}, {Donovan}, {Ebeling}, {Shandera}, \&
  {Wright}}]{mantz15}
{Mantz}, A.~B., {von der Linden}, A., {Allen}, S.~W., {et~al.} 2015, \mnras,
  446, 2205

\bibitem[{{Marulli} {et~al.}(2018){Marulli}, {Veropalumbo}, {Sereno},
  {Moscardini}, {Pacaud}, {Pierre}, {Plionis}, {Cappi}, {Adami}, {Alis},
  {Altieri}, {Birkinshaw}, {Ettori}, {Faccioli}, {Gastaldello}, {Koulouridis},
  {Lidman}, {Le F\ `evre}, {Maurogordato}, {Poggianti}, {Pompei}, {Sadibekova},
  \& {Valtchanov}}]{xxlpaperXVI}
{Marulli}, F., {Veropalumbo}, A., {Sereno}, M., {et~al.} 2018, \aap, in press,
  DOI 10.1051/0004-6361/201833238 (XXL paper XVI), arXiv eprints
  [\eprint[arXiv]{1807.04760}]

\bibitem[{{Pacaud} {et~al.}(2016){Pacaud}, {Clerc}, {Giles}, {Adami},
  {Sadibekova}, {Pierre}, {Maughan}, {Lieu}, {Le F{\`e}vre}, {Alis}, {Altieri},
  {Ardila}, {Baldry}, {Benoist}, {Birkinshaw}, {Chiappetti},
  {D{\'e}mocl{\`e}s}, {Eckert}, {Evrard}, {Faccioli}, {Gastaldello}, {Guennou},
  {Horellou}, {Iovino}, {Koulouridis}, {Le Brun}, {Lidman}, {Liske},
  {Maurogordato}, {Menanteau}, {Owers}, {Poggianti}, {Pomar{\`e}de}, {Pompei},
  {Ponman}, {Rapetti}, {Reiprich}, {Smith}, {Tuffs}, {Valageas}, {Valtchanov},
  {Willis}, \& {Ziparo}}]{xxlpaperII}
{Pacaud}, F., {Clerc}, N., {Giles}, P.~A., {et~al.} 2016, \aap, 592, A2, (XXL
  Paper II)

\bibitem[{{Pacaud} {et~al.}(2006){Pacaud}, {Pierre}, {Refregier}, {Gueguen},
  {Starck}, {Valtchanov}, {Read}, {Altieri}, {Chiappetti}, {Gandhi}, {Garcet},
  {Gosset}, {Ponman}, \& {Surdej}}]{pacaud06}
{Pacaud}, F., {Pierre}, M., {Refregier}, A., {et~al.} 2006, \mnras, 372, 578

\bibitem[{{Pierre} {et~al.}(2016){Pierre}, {Pacaud}, {Adami}, {Alis},
  {Altieri}, {Baran}, {Benoist}, {Birkinshaw}, {Bongiorno}, {Bremer}, {Brusa},
  {Butler}, {Ciliegi}, {Chiappetti}, {Clerc}, {Corasaniti}, {Coupon}, {De
  Breuck}, {Democles}, {Desai}, {Delhaize}, {Devriendt}, {Dubois}, {Eckert},
  {Elyiv}, {Ettori}, {Evrard}, {Faccioli}, {Farahi}, {Ferrari}, {Finet},
  {Fotopoulou}, {Fourmanoit}, {Gandhi}, {Gastaldello}, {Gastaud},
  {Georgantopoulos}, {Giles}, {Guennou}, {Guglielmo}, {Horellou}, {Husband},
  {Huynh}, {Iovino}, {Kilbinger}, {Koulouridis}, {Lavoie}, {Le Brun}, {Le
  Fevre}, {Lidman}, {Lieu}, {Lin}, {Mantz}, {Maughan}, {Maurogordato},
  {McCarthy}, {McGee}, {Melin}, {Melnyk}, {Menanteau}, {Novak}, {Paltani},
  {Plionis}, {Poggianti}, {Pomarede}, {Pompei}, {Ponman}, {Ramos-Ceja},
  {Ranalli}, {Rapetti}, {Raychaudury}, {Reiprich}, {Rottgering}, {Rozo},
  {Rykoff}, {Sadibekova}, {Santos}, {Sauvageot}, {Schimd}, {Sereno}, {Smith},
  {Smol{\v c}i{\'c}}, {Snowden}, {Spergel}, {Stanford}, {Surdej}, {Valageas},
  {Valotti}, {Valtchanov}, {Vignali}, {Willis}, \& {Ziparo}}]{xxlpaperI}
{Pierre}, M., {Pacaud}, F., {Adami}, C., {et~al.} 2016, \aap, 592, A1, (XXL
  Paper I)

\bibitem[{{Pierre} {et~al.}(2011){Pierre}, {Pacaud}, {Juin}, {Melin},
  {Valageas}, {Clerc}, \& {Corasaniti}}]{pierre11}
{Pierre}, M., {Pacaud}, F., {Juin}, J.~B., {et~al.} 2011, \mnras, 414, 1732

\bibitem[{{Planck Collaboration} {et~al.}(2016{\natexlab{a}}){Planck
  Collaboration}, {Ade}, {Aghanim}, {Arnaud}, {Ashdown}, {Aumont},
  {Baccigalupi}, {Banday}, {Barreiro}, {Bartlett}, \& et~al.}]{planck2015XIII}
{Planck Collaboration}, {Ade}, P.~A.~R., {Aghanim}, N., {et~al.}
  2016{\natexlab{a}}, \aap, 594, A13

\bibitem[{{Planck Collaboration} {et~al.}(2016{\natexlab{b}}){Planck
  Collaboration}, {Ade}, {Aghanim}, {Arnaud}, {Ashdown}, {Aumont},
  {Baccigalupi}, {Banday}, {Barreiro}, {Bartlett}, \& et~al.}]{planck2015XV}
{Planck Collaboration}, {Ade}, P.~A.~R., {Aghanim}, N., {et~al.}
  2016{\natexlab{b}}, \aap, 594, A15

\bibitem[{{Planck Collaboration} {et~al.}(2016{\natexlab{c}}){Planck
  Collaboration}, {Ade}, {Aghanim}, {Arnaud}, {Ashdown}, {Aumont},
  {Baccigalupi}, {Banday}, {Barreiro}, {Bartlett}, \& et~al.}]{planck2015XXIV}
{Planck Collaboration}, {Ade}, P.~A.~R., {Aghanim}, N., {et~al.}
  2016{\natexlab{c}}, \aap, 594, A24

\bibitem[{{Planck Collaboration} {et~al.}(2018){Planck Collaboration},
  {Aghanim}, {Akrami}, {Ashdown}, {Aumont}, {Baccigalupi}, {Ballardini},
  {Banday}, {Barreiro}, {Bartolo}, {Basak}, {Battye}, {Benabed}, {Bernard},
  {Bersanelli}, {Bielewicz}, {Bock}, {Bond}, {Borrill}, {Bouchet}, {Boulanger},
  {Bucher}, {Burigana}, {Butler}, {Calabrese}, {Cardoso}, {Carron},
  {Challinor}, {Chiang}, {Chluba}, {Colombo}, {Combet}, {Contreras}, {Crill},
  {Cuttaia}, {de Bernardis}, {de Zotti}, {Delabrouille}, {Delouis}, {Di
  Valentino}, {Diego}, {Dor{\'e}}, {Douspis}, {Ducout}, {Dupac}, {Dusini},
  {Efstathiou}, {Elsner}, {En{\ss}lin}, {Eriksen}, {Fantaye}, {Farhang},
  {Fergusson}, {Fernandez-Cobos}, {Finelli}, {Forastieri}, {Frailis},
  {Franceschi}, {Frolov}, {Galeotta}, {Galli}, {Ganga}, {G{\'e}nova-Santos},
  {Gerbino}, {Ghosh}, {Gonz{\'a}lez-Nuevo}, {G{\'o}rski}, {Gratton},
  {Gruppuso}, {Gudmundsson}, {Hamann}, {Handley}, {Herranz}, {Hivon}, {Huang},
  {Jaffe}, {Jones}, {Karakci}, {Keih{\"a}nen}, {Keskitalo}, {Kiiveri}, {Kim},
  {Kisner}, {Knox}, {Krachmalnicoff}, {Kunz}, {Kurki-Suonio}, {Lagache},
  {Lamarre}, {Lasenby}, {Lattanzi}, {Lawrence}, {Le Jeune}, {Lemos},
  {Lesgourgues}, {Levrier}, {Lewis}, {Liguori}, {Lilje}, {Lilley}, {Lindholm},
  {L{\'o}pez-Caniego}, {Lubin}, {Ma}, {Mac{\'{\i}}as-P{\'e}rez}, {Maggio},
  {Maino}, {Mandolesi}, {Mangilli}, {Marcos-Caballero}, {Maris}, {Martin},
  {Martinelli}, {Mart{\'{\i}}nez-Gonz{\'a}lez}, {Matarrese}, {Mauri}, {McEwen},
  {Meinhold}, {Melchiorri}, {Mennella}, {Migliaccio}, {Millea}, {Mitra},
  {Miville-Desch{\^e}nes}, {Molinari}, {Montier}, {Morgante}, {Moss}, {Natoli},
  {N{\o}rgaard-Nielsen}, {Pagano}, {Paoletti}, {Partridge}, {Patanchon},
  {Peiris}, {Perrotta}, {Pettorino}, {Piacentini}, {Polastri}, {Polenta},
  {Puget}, {Rachen}, {Reinecke}, {Remazeilles}, {Renzi}, {Rocha}, {Rosset},
  {Roudier}, {Rubi{\~n}o-Mart{\'{\i}}n}, {Ruiz-Granados}, {Salvati}, {Sandri},
  {Savelainen}, {Scott}, {Shellard}, {Sirignano}, {Sirri}, {Spencer},
  {Sunyaev}, {Suur-Uski}, {Tauber}, {Tavagnacco}, {Tenti}, {Toffolatti},
  {Tomasi}, {Trombetti}, {Valenziano}, {Valiviita}, {Van Tent}, {Vibert},
  {Vielva}, {Villa}, {Vittorio}, {Wandelt}, {Wehus}, {White}, {White},
  {Zacchei}, \& {Zonca}}]{planck2018VI}
{Planck Collaboration}, {Aghanim}, N., {Akrami}, Y., {et~al.} 2018, ArXiv
  e-prints [\eprint[arXiv]{1807.06209}]

\bibitem[{{Planck Collaboration} {et~al.}(2017){Planck Collaboration},
  {Aghanim}, {Akrami}, {Ashdown}, {Aumont}, {Baccigalupi}, {Ballardini},
  {Banday}, {Barreiro}, {Bartolo}, {Basak}, {Benabed}, {Bersanelli},
  {Bielewicz}, {Bonaldi}, {Bonavera}, {Bond}, {Borrill}, {Bouchet}, {Burigana},
  {Calabrese}, {Cardoso}, {Challinor}, {Chiang}, {Colombo}, {Combet}, {Crill},
  {Curto}, {Cuttaia}, {de Bernardis}, {de Rosa}, {de Zotti}, {Delabrouille},
  {Di Valentino}, {Dickinson}, {Diego}, {Dor{\'e}}, {Ducout}, {Dupac},
  {Dusini}, {Efstathiou}, {Elsner}, {En{\ss}lin}, {Eriksen}, {Fantaye},
  {Finelli}, {Forastieri}, {Frailis}, {Franceschi}, {Frolov}, {Galeotta},
  {Galli}, {Ganga}, {G{\'e}nova-Santos}, {Gerbino}, {Gonz{\'a}lez-Nuevo},
  {G{\'o}rski}, {Gratton}, {Gruppuso}, {Gudmundsson}, {Herranz}, {Hivon},
  {Huang}, {Jaffe}, {Jones}, {Keih{\"a}nen}, {Keskitalo}, {Kiiveri}, {Kim},
  {Kisner}, {Knox}, {Krachmalnicoff}, {Kunz}, {Kurki-Suonio}, {Lagache},
  {Lamarre}, {Lasenby}, {Lattanzi}, {Lawrence}, {Le Jeune}, {Levrier}, {Lewis},
  {Liguori}, {Lilje}, {Lilley}, {Lindholm}, {L{\'o}pez-Caniego}, {Lubin}, {Ma},
  {Mac{\'{\i}}as-P{\'e}rez}, {Maggio}, {Maino}, {Mandolesi}, {Mangilli},
  {Maris}, {Martin}, {Mart{\'{\i}}nez-Gonz{\'a}lez}, {Matarrese}, {Mauri},
  {McEwen}, {Meinhold}, {Mennella}, {Migliaccio}, {Millea},
  {Miville-Desch{\^e}nes}, {Molinari}, {Moneti}, {Montier}, {Morgante}, {Moss},
  {Narimani}, {Natoli}, {Oxborrow}, {Pagano}, {Paoletti}, {Partridge},
  {Patanchon}, {Patrizii}, {Pettorino}, {Piacentini}, {Polastri}, {Polenta},
  {Puget}, {Rachen}, {Racine}, {Reinecke}, {Remazeilles}, {Renzi}, {Rocha},
  {Rossetti}, {Roudier}, {Rubi{\~n}o-Mart{\'{\i}}n}, {Ruiz-Granados},
  {Salvati}, {Sandri}, {Savelainen}, {Scott}, {Sirignano}, {Sirri}, {Stanco},
  {Suur-Uski}, {Tauber}, {Tavagnacco}, {Tenti}, {Toffolatti}, {Tomasi},
  {Tristram}, {Trombetti}, {Valiviita}, {Van Tent}, {Vielva}, {Villa},
  {Vittorio}, {Wandelt}, {Wehus}, {White}, {Zacchei}, \&
  {Zonca}}]{planckIntermLI}
{Planck Collaboration}, {Aghanim}, N., {Akrami}, Y., {et~al.} 2017, \aap, 607,
  A95

\bibitem[{{Planck Collaboration} {et~al.}(2016{\natexlab{d}}){Planck
  Collaboration}, {Aghanim}, {Arnaud}, {Ashdown}, {Aumont}, {Baccigalupi},
  {Banday}, {Barreiro}, {Bartlett}, {Bartolo}, \& et~al.}]{planck2015XI}
{Planck Collaboration}, {Aghanim}, N., {Arnaud}, M., {et~al.}
  2016{\natexlab{d}}, \aap, 594, A11

\bibitem[{{Planck Collaboration} {et~al.}(2016{\natexlab{e}}){Planck
  Collaboration}, {Aghanim}, {Ashdown}, {Aumont}, {Baccigalupi}, {Ballardini},
  {Banday}, {Barreiro}, {Bartolo}, {Basak}, {Battye}, {Benabed}, {Bernard},
  {Bersanelli}, {Bielewicz}, {Bock}, {Bonaldi}, {Bonavera}, {Bond}, {Borrill},
  {Bouchet}, {Boulanger}, {Bucher}, {Burigana}, {Butler}, {Calabrese},
  {Cardoso}, {Carron}, {Challinor}, {Chiang}, {Colombo}, {Combet}, {Comis},
  {Coulais}, {Crill}, {Curto}, {Cuttaia}, {Davis}, {de Bernardis}, {de Rosa},
  {de Zotti}, {Delabrouille}, {Delouis}, {Di Valentino}, {Dickinson}, {Diego},
  {Dor{\'e}}, {Douspis}, {Ducout}, {Dupac}, {Efstathiou}, {Elsner},
  {En{\ss}lin}, {Eriksen}, {Falgarone}, {Fantaye}, {Finelli}, {Forastieri},
  {Frailis}, {Fraisse}, {Franceschi}, {Frolov}, {Galeotta}, {Galli}, {Ganga},
  {G{\'e}nova-Santos}, {Gerbino}, {Ghosh}, {Gonz{\'a}lez-Nuevo}, {G{\'o}rski},
  {Gratton}, {Gruppuso}, {Gudmundsson}, {Hansen}, {Helou},
  {Henrot-Versill{\'e}}, {Herranz}, {Hivon}, {Huang}, {Ili{\'c}}, {Jaffe},
  {Jones}, {Keih{\"a}nen}, {Keskitalo}, {Kisner}, {Knox}, {Krachmalnicoff},
  {Kunz}, {Kurki-Suonio}, {Lagache}, {Lamarre}, {Langer}, {Lasenby},
  {Lattanzi}, {Lawrence}, {Le Jeune}, {Leahy}, {Levrier}, {Liguori}, {Lilje},
  {L{\'o}pez-Caniego}, {Ma}, {Mac{\'{\i}}as-P{\'e}rez}, {Maggio}, {Mangilli},
  {Maris}, {Martin}, {Mart{\'{\i}}nez-Gonz{\'a}lez}, {Matarrese}, {Mauri},
  {McEwen}, {Meinhold}, {Melchiorri}, {Mennella}, {Migliaccio},
  {Miville-Desch{\^e}nes}, {Molinari}, {Moneti}, {Montier}, {Morgante}, {Moss},
  {Mottet}, {Naselsky}, {Natoli}, {Oxborrow}, {Pagano}, {Paoletti},
  {Partridge}, {Patanchon}, {Patrizii}, {Perdereau}, {Perotto}, {Pettorino},
  {Piacentini}, {Plaszczynski}, {Polastri}, {Polenta}, {Puget}, {Rachen},
  {Racine}, {Reinecke}, {Remazeilles}, {Renzi}, {Rocha}, {Rossetti}, {Roudier},
  {Rubi{\~n}o-Mart{\'{\i}}n}, {Ruiz-Granados}, {Salvati}, {Sandri},
  {Savelainen}, {Scott}, {Sirri}, {Sunyaev}, {Suur-Uski}, {Tauber}, {Tenti},
  {Toffolatti}, {Tomasi}, {Tristram}, {Trombetti}, {Valiviita}, {Van Tent},
  {Vibert}, {Vielva}, {Villa}, {Vittorio}, {Wandelt}, {Watson}, {Wehus},
  {White}, {Zacchei}, \& {Zonca}}]{planckIntermXLVI}
{Planck Collaboration}, {Aghanim}, N., {Ashdown}, M., {et~al.}
  2016{\natexlab{e}}, \aap, 596, A107

\bibitem[{{Riess} {et~al.}(2018){Riess}, {Casertano}, {Yuan}, {Macri},
  {Bucciarelli}, {Lattanzi}, {MacKenty}, {Bowers}, {Zheng}, {Filippenko},
  {Huang}, \& {Anderson}}]{riess18}
{Riess}, A.~G., {Casertano}, S., {Yuan}, W., {et~al.} 2018, \apj, 861, 126

\bibitem[{{Rozo} {et~al.}(2014){Rozo}, {Rykoff}, {Bartlett}, \&
  {Evrard}}]{rozo14}
{Rozo}, E., {Rykoff}, E.~S., {Bartlett}, J.~G., \& {Evrard}, A. 2014, \mnras,
  438, 49

\bibitem[{{Salvati} {et~al.}(2018){Salvati}, {Douspis}, \&
  {Aghanim}}]{salvati18}
{Salvati}, L., {Douspis}, M., \& {Aghanim}, N. 2018, \aap, 614, A13

\bibitem[{{Sereno} {et~al.}(2017){Sereno}, {Covone}, {Izzo}, {Ettori},
  {Coupon}, \& {Lieu}}]{sereno17}
{Sereno}, M., {Covone}, G., {Izzo}, L., {et~al.} 2017, \mnras, 472, 1946

\bibitem[{{Sereno} \& {Ettori}(2015)}]{sereno15}
{Sereno}, M. \& {Ettori}, S. 2015, \mnras, 450, 3633

\bibitem[{{Smith} {et~al.}(2001){Smith}, {Brickhouse}, {Liedahl}, \&
  {Raymond}}]{smith01}
{Smith}, R.~K., {Brickhouse}, N.~S., {Liedahl}, D.~A., \& {Raymond}, J.~C.
  2001, \apjl, 556, L91

\bibitem[{{Stanek} {et~al.}(2009){Stanek}, {Rudd}, \& {Evrard}}]{stanek09}
{Stanek}, R., {Rudd}, D., \& {Evrard}, A.~E. 2009, \mnras, 394, L11

\bibitem[{{Tinker} {et~al.}(2008){Tinker}, {Kravtsov}, {Klypin}, {Abazajian},
  {Warren}, {Yepes}, {Gottl{\"o}ber}, \& {Holz}}]{tinker08}
{Tinker}, J., {Kravtsov}, A.~V., {Klypin}, A., {et~al.} 2008, \apj, 688, 709

\bibitem[{{Troxel} {et~al.}(2017){Troxel}, {MacCrann}, {Zuntz}, {Eifler},
  {Krause}, {Dodelson}, {Gruen}, {Blazek}, {Friedrich}, {Samuroff}, {Prat},
  {Secco}, {Davis}, {Fert{\'e}}, {DeRose}, {Alarcon}, {Amara}, {Baxter},
  {Becker}, {Bernstein}, {Bridle}, {Cawthon}, {Chang}, {Choi}, {De Vicente},
  {Drlica-Wagner}, {Elvin-Poole}, {Frieman}, {Gatti}, {Hartley}, {Honscheid},
  {Hoyle}, {Huff}, {Huterer}, {Jain}, {Jarvis}, {Kacprzak}, {Kirk}, {Kokron},
  {Krawiec}, {Lahav}, {Liddle}, {Peacock}, {Rau}, {Refregier}, {Rollins},
  {Rozo}, {Rykoff}, {S{\'a}nchez}, {Sevilla-Noarbe}, {Sheldon}, {Stebbins},
  {Varga}, {Vielzeuf}, {Wang}, {Wechsler}, {Yanny}, {Abbott}, {Abdalla},
  {Allam}, {Annis}, {Bechtol}, {Benoit-L{\'e}vy}, {Bertin}, {Brooks},
  {Buckley-Geer}, {Burke}, {Carnero Rosell}, {Carrasco Kind}, {Carretero},
  {Castander}, {Crocce}, {Cunha}, {D'Andrea}, {da Costa}, {DePoy}, {Desai},
  {Diehl}, {Dietrich}, {Doel}, {Fernandez}, {Flaugher}, {Fosalba},
  {Garc{\'{\i}}a-Bellido}, {Gaztanaga}, {Gerdes}, {Giannantonio}, {Goldstein},
  {Gruendl}, {Gschwend}, {Gutierrez}, {James}, {Jeltema}, {Johnson}, {Johnson},
  {Kent}, {Kuehn}, {Kuhlmann}, {Kuropatkin}, {Li}, {Lima}, {Lin}, {Maia},
  {March}, {Marshall}, {Martini}, {Melchior}, {Menanteau}, {Miquel}, {Mohr},
  {Neilsen}, {Nichol}, {Nord}, {Petravick}, {Plazas}, {Romer}, {Roodman},
  {Sako}, {Sanchez}, {Scarpine}, {Schindler}, {Schubnell}, {Smith}, {Smith},
  {Soares-Santos}, {Sobreira}, {Suchyta}, {Swanson}, {Tarle}, {Thomas},
  {Tucker}, {Vikram}, {Walker}, {Weller}, \& {Zhang}}]{troxel18}
{Troxel}, M.~A., {MacCrann}, N., {Zuntz}, J., {et~al.} 2017, ArXiv e-prints
  [\eprint[arXiv]{1708.01538}]

\bibitem[{{Valageas} {et~al.}(2011){Valageas}, {Clerc}, {Pacaud}, \&
  {Pierre}}]{valageas11}
{Valageas}, P., {Clerc}, N., {Pacaud}, F., \& {Pierre}, M. 2011, \aap, 536, A95

\bibitem[{{Valotti} {et~al.}(2018){Valotti}, {Pierre}, {Farahi}, {Evrard},
  {Faccioli}, {Sauvageot}, {Clerc}, \& {Pacaud}}]{valotti18}
{Valotti}, A., {Pierre}, M., {Farahi}, A., {et~al.} 2018, \aap, 614, A72

\bibitem[{{Velliscig} {et~al.}(2014){Velliscig}, {van Daalen}, {Schaye},
  {McCarthy}, {Cacciato}, {Le Brun}, \& {Dalla Vecchia}}]{velliscig14}
{Velliscig}, M., {van Daalen}, M.~P., {Schaye}, J., {et~al.} 2014, \mnras, 442,
  2641

\bibitem[{{von der Linden} {et~al.}(2014){von der Linden}, {Mantz}, {Allen},
  {Applegate}, {Kelly}, {Morris}, {Wright}, {Allen}, {Burchat}, {Burke},
  {Donovan}, \& {Ebeling}}]{vonderlinden14}
{von der Linden}, A., {Mantz}, A., {Allen}, S.~W., {et~al.} 2014, \mnras, 443,
  1973

\bibitem[{{Watson} {et~al.}(2013){Watson}, {Iliev}, {D'Aloisio},
  {et~al.}}]{watson13}
{Watson}, W.~A., {Iliev}, I.~T., {D'Aloisio}, A., {et~al.} 2013, \mnras, 433,
  1230

\end{thebibliography}

\begin{appendix}

\section{ C1 cluster likelihood model}
\label{append:Like}

In this paper, we obtain cosmological constraints from the density and redshift distribution 
of the C1 galaxy cluster sample. Our analysis relies on the likelihood model described in \citetalias{xxlpaperII}, which we  summarise here.

The first step in the calculation is to derive the density of galaxy clusters in a given cosmology as a function of their ICM properties.
Starting from the differential mass function expressed in terms of redshift ($z$) and sky area ($\Omega$),  $\frac{\mathrm{d}n\left(M_\mathrm{500},z\right)}{\mathrm{d}M_\mathrm{500}\,\mathrm{d}\Omega\,\mathrm{d}z}$, 
we use the $M_\mathrm{500,WL}$--$T_{300\mathrm{kpc}}$ and $L^\mathrm{XXL}_\mathrm{500}$--$T_{300\mathrm{kpc}}$ 
scaling relations to derive an equivalent temperature function, without including the scatter, 
and disperse it over a luminosity distribution
\begin{equation}
  \frac{\mathrm{d}n\left(L,T,z\right)}{ \mathrm{d}z\,\mathrm{d}L\,\mathrm{d}T} =
    \frac{\mathrm{d}n\left(M,z\right)}{\mathrm{d}M\,\mathrm{d}\Omega\,\mathrm{d}z}\frac{\mathrm{d}M(T,z)}{\mathrm{d}T} \mathcal{LN}\left[L\,|\,\hat{L}\left(T,z\right),\sigma_{LT}\right],
    \label{eq:dnOverdLdT}
\end{equation}
where $T$ and  $\hat{L}$ are the average temperature and [0.5--2.0] keV luminosities at a given 
mass obtained from the scaling relations of \autoref{scalrel}, and 
$\mathcal{LN}\left[ L\,|\,\hat{L},\,\sigma\right]$ is a log-normal distribution of mean 
$\hat{L}$ and scatter $\sigma$.

The combination of this distribution with the survey effective sky coverage allows us to 
derive the cluster redshift density for our analysis. The selection function of the XXL C1 
clusters in terms of raw observables is discussed in section~5 of \citetalias{xxlpaperII}
and depends on the source total count rate ($CR_\mathrm{\infty}$) and the angular core radius 
$\theta_\mathrm{c}$ of a $\beta$-model with $\beta$=2/3.  
The corresponding sky coverage $\Omega_\mathrm{S}(CR_\mathrm{\infty},\theta_\mathrm{c})$ must be recast as a function of cluster physical properties. First, we derive a core radius from the characteristic size of the clusters, $r_\mathrm{500}=\left[\frac{3 M_\mathrm{500}(T)}{4\pi\times500\rho_\mathrm{c}}\right]^{1/3}$, and the size scaling relation of \autoref{scalrel}, which can be expressed through the constant parameter $x_\mathrm{c} = r_\mathrm{c}/r_\mathrm{500}$. 
Second, we use an APEC thermal model \citep[version 3.0.9]{smith01} with a metallicity set to 0.3 times the solar value to estimate the source count rate within $r_\mathrm{500}$, based on the cluster luminosity, $L^\mathrm{XXL}_\mathrm{500}$, and temperature.
In addition, an extrapolation factor from $r_\mathrm{500}$ to infinity, $f_\mathrm{\infty}$, is computed from the assumed $\beta$-model profile. This results in an  effective sky coverage,  
\begin{equation}
   \Omega_\mathrm{S}\left(L,T,z\right) = \Omega_\mathrm{S}\left(f_\mathrm{\infty}CR_\mathrm{500}\left[L,T,z\right],r_\mathrm{c}\left[T,z\right]/d_\mathrm{A}\left[z\right]\right),
   \label{eq:OmegaLTz}
\end{equation}
and a final redshift distribution for the model,
\begin{equation}
   \frac{\mathrm{d}n}{\mathrm{d}z}(z) = \int_0^\infty\int_0^\infty  \Omega_\mathrm{S}\left(L,T,z\right)\frac{\mathrm{d}n\left(L,T,z\right)}{ \mathrm{d}z\,\mathrm{d}L\,\mathrm{d}T} \mathrm{d}L\mathrm{d}T
   \label{eq:dNdZ}.
\end{equation}
The total number of clusters predicted by the model, $N_\mathrm{tot}$ follows from a simple redshift integration.

To infer model parameters ($\mathcal{P}$) from the properties of the C1 clusters,
we make use of a very generic unbinned likelihood model, in which we separate the information
on the number of detected clusters, $N_\mathrm{det}$, from their redshift distribution,
\begin{equation}
        \mathcal{L}(\mathcal{P}) = P(N_\mathrm{det}|N_\mathrm{tot},\mathcal{P}) \prod_{i=1}^{N_{det}} \left[\frac{1}{N_\mathrm{tot}} \frac{\mathrm{d}n}{\mathrm{d}z}(z_i) \right],
    \label{eq:likeForm}
\end{equation}
where $P(N_\mathrm{det}|N_\mathrm{tot},\mathcal{P})$ describes the probability of observing
$N_\mathrm{det}$ clusters in a given cosmological model. A standard choice for this probability 
would be to use a Poisson law of parameter $N_\mathrm{tot}$, but we opted for a more complicated 
distribution in order to account for the significant cosmic variance within the XXL fields. 
We estimate this variance, $\sigma_\mathrm{v}^2$, with the formalism presented in \citet{valageas11}.
For cosmological models which provide a good description of the XXL cluster population, 
this super-sample variance term amounts to $\sim$30\% of the sample Poisson variance.
The combined distribution from shot noise and cosmic variance is modeled as 
\begin{equation}
    P(N_\mathrm{det}|\mathcal{P}) = \int \mathrm{Po}(N_\mathrm{det}|N_\mathrm{loc})\,\mathcal{LN}\left[N_\mathrm{loc}|{\langle}N_\mathrm{det}{\rangle},\sigma_\mathrm{v}\right]\,\mathrm{d}N_\mathrm{loc}
    \label{eq:CosVar}
,\end{equation}
where the local density, $N_\mathrm{loc}$, is generated from a log-normal distribution $\mathcal{LN}$ of
mean ${\langle}N_\mathrm{det}{\rangle}$ and sample variance $\sigma_\mathrm{v}^2$, and $N_\mathrm{loc}$ is then subjected 
to additional shot noise through the Poisson law $\mathrm{Po(x|\lambda)}$. 

For all the results presented in this article, we sample this likelihood using a Metropolis 
Markov chain Monte Carlo (MCMC) algorithm, combined with the priors listed in \autoref{sec:Methods}
and non-informative priors on all other parameters. We run four chains in parallel, excluding a 20\% burn-in phase and monitor the convergence with the Gelman--Rubin diagnostic \citep{gelman92}.
The chains are stopped when they reach a convergence of $R-1<0.03$.
As mentioned in the caption of \autoref{scalrel}, two scaling relation parameters are left free 
in the process, the normalisation of the $M_\mathrm{500,WL}$--$T_\mathrm{300kpc}$ relation and the redshift evolution of the $L^\mathrm{XXL}_\mathrm{500}$--$T_\mathrm{300kpc}$ relation. 
These are constrained within the fits by priors derived from earlier XXL scaling relation analyses \citepalias{xxlpaperIV,xxlpaperXX}. 

Finally, the combination of the C1 cluster results with other cosmological probes (Planck, KIDS) 
relies on importance sampling of the respective chains based on the C1 likelihood, without 
any prior. In the specific case of the KIDS survey, \citet{hildebrandt17} already applies 
top-hat priors that are similar   to ours on $0.019<\Omega_b h^2<0.026$ and $0.064<h<0.82$. 
However, the prior on $0.01<\Omega_m<0.99$ is extremely wide and therefore there is no direct prior applied in the $\Omega_m$--$\sigma_8$ plane for the combined XXL+KIDS constraints.

\section{Cosmological constraints from CMB observations}
\label{append:CMB}

Recently, \citet{planckIntermXLVI} has described new calibration and data processing methods 
which  improve the control of systematics in the CMB polarisation maps obtained with
the Planck HFI instrument. This has a significant impact on the determination of the optical
depth to reionisation, $\tau$, which is almost fully degenerate with the amplitude of matter 
fluctuation in the temperature power spectrum, but shows distinct signatures on the large-scale polarisation signal. As a result, the authors obtained unprecedented constraints
on this parameter, $\tau=0.055\pm0.009$, which is systematically lower than all previous estimates (e.g. $\tau=0.066\pm0.016$ in \citealt{planck2015XIII}). Such a decrease 
in the optical depth directly translates to a lower amplitude of the matter fluctuations at 
the epoch of recombination in order to fit the CMB data. It was immediately recognised as a possible route to soften the tensions between the preferred Planck cosmological model and low-redshift probes of the large-scale structures 
\citep[e.g.][]{salvati18}. 

A meaningful comparison with the XXL cluster sample therefore requires a new CMB analysis that accounts for the updated constraints on $\tau$. Unfortunately, at the time when the core work of this article was being performed, these results had not yet been released by the Planck collaboration, nor were the improved polarisation maps and power spectra available.
Instead, we had to use the public Planck likelihood codes \citep{planck2015XI} to generate updated sets of cosmological constraints, based on the Planck 2015 dataset.  
In doing so, we only account for the temperature power spectrum (TT) at both high and low multipole values ($\ell$), we ignored any polarisation constraints, but replaced the low-$\ell$ polarisation likelihood by a Gaussian prior on $\tau=0.055\pm0.009$ mimicking the measurement obtained by \cite{planckIntermXLVI}.  

In addition to the temperature and polarisation power spectra, the Planck Collaboration also released a reconstructed map of the lensing potential distorting the CMB, as well as its power spectrum \citep{planck2015XV}. The latter can also be used to constrain cosmological parameters based on the large-scale structures at intermediate redshifts 
(with maximum contribution from $z\sim2-3$). We include the official Planck likelihood for the power spectrum of the lensing potential in our reanalysis. Intuitively, excluding the lensing constraints from the analysis might seem to better decouple probes of early large-scale structures (the primary CMB) from late time tracers (the XXL clusters); however, this  is actually not the case.
The same lensing effects are indeed already included in the analysis of the temperature power spectrum and, as shown in \cite{planckIntermLI}, play a major role in the derivation of a high $\sigma_8$ value from Planck: high matter fluctuations are favoured to explain the significant smoothing of high-$\ell$ acoustic peaks, a natural consequence of CMB lensing. However, the direct modelling of the lensing potential power spectrum does not require such high fluctuations and adding it to the analysis provides a more balanced view of the constraints originating from CMB lensing.

\begin{figure*}
\includegraphics[width=\textwidth]{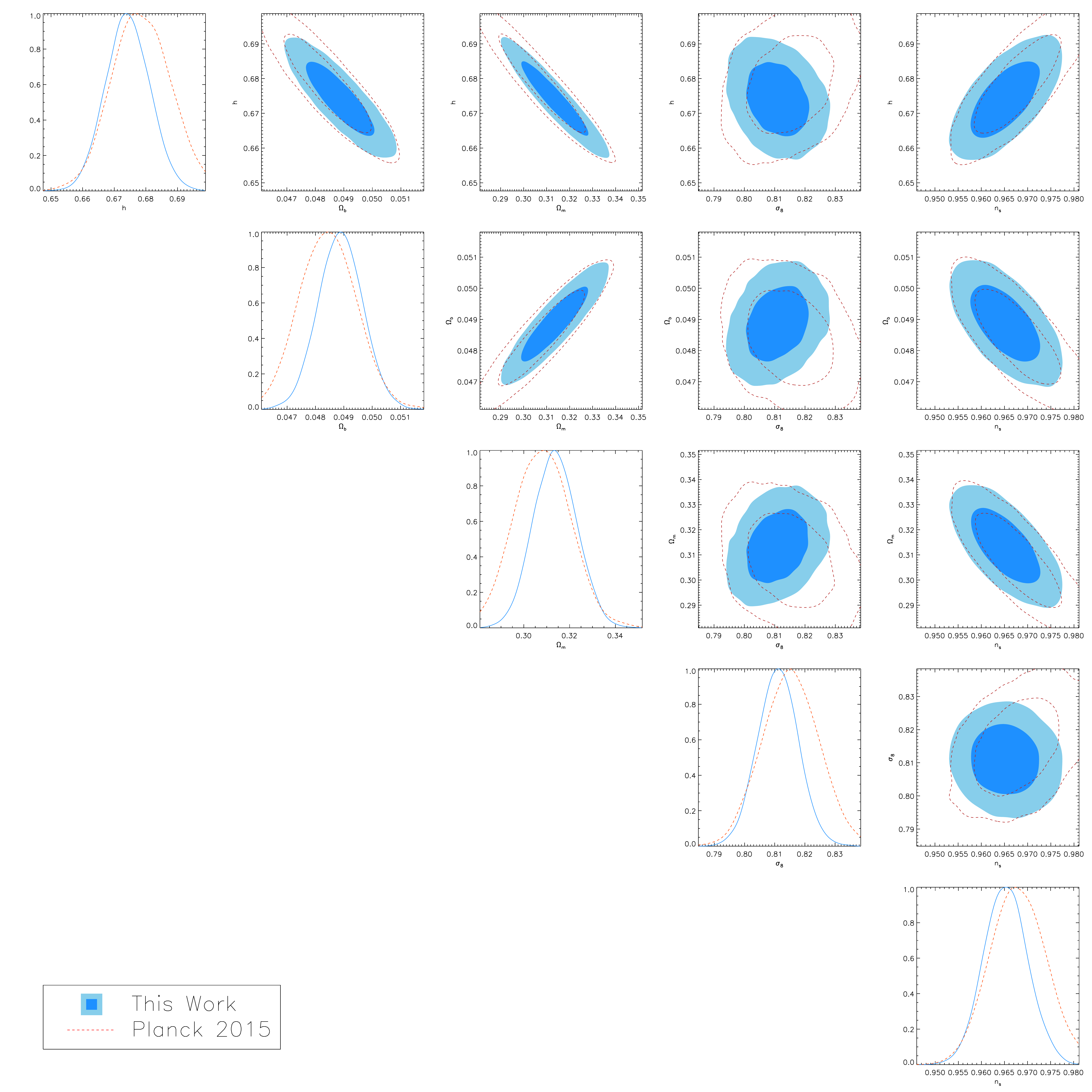}
\caption{Comparison of the cosmological analysis of the Planck CMB products used in this 
paper with the original constraints of \citet{planck2015XIII} in a flat $\mathrm{{\Lambda}CDM}$
universe (1$\sigma$ and 2$\sigma$ contours).}
\label{fig:CosParamLambda}
\end{figure*}

\begin{figure*}
\includegraphics[width=\textwidth]{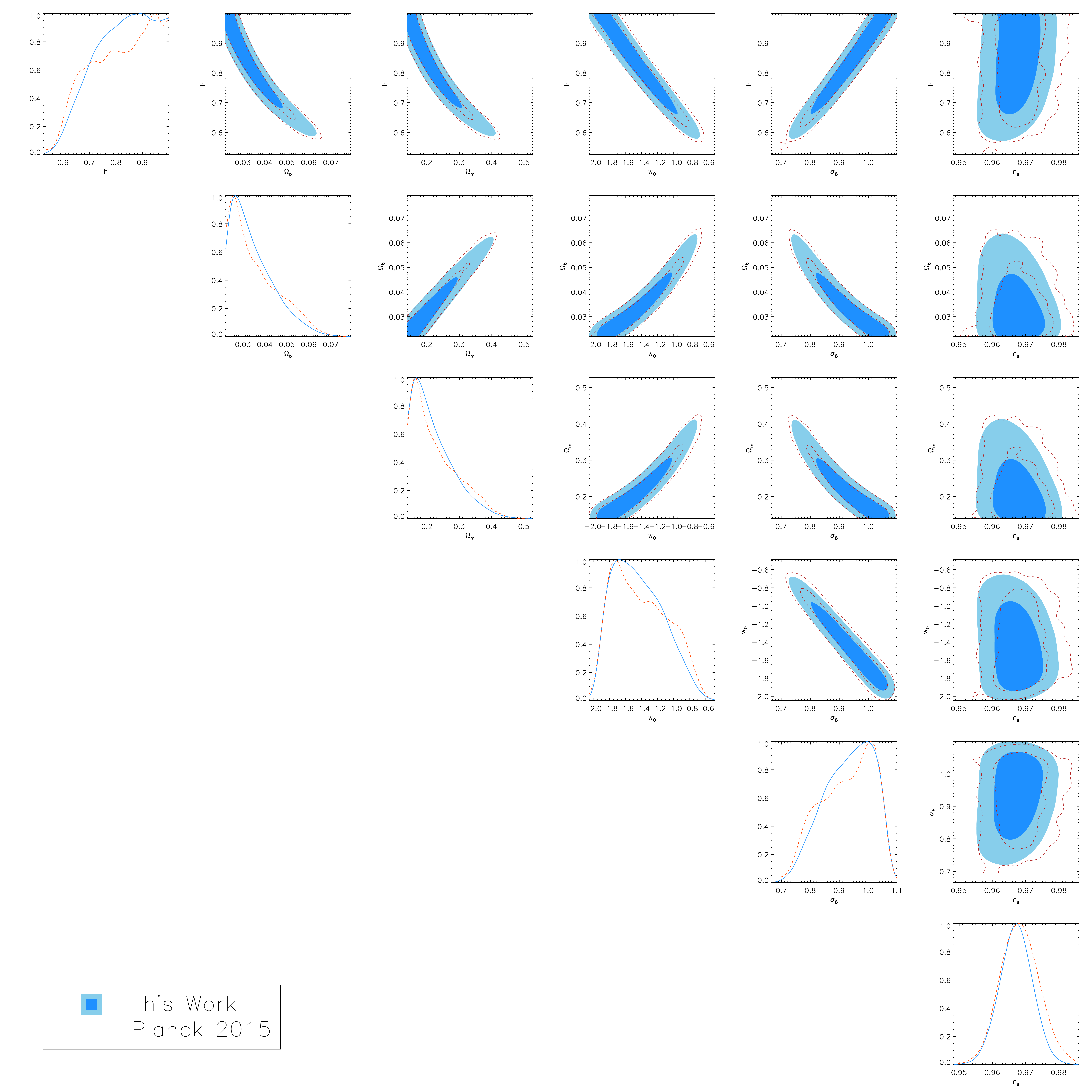}
\caption{Comparison of the cosmological analysis of the Planck CMB products used in this 
paper with the original constraints of \citet{planck2015XIII} in a flat $\mathrm{wCDM}$
universe (1$\sigma$ and 2$\sigma$ contours).}
\label{fig:CosParamW}
\end{figure*}

\autoref{fig:CosParamLambda} and \ref{fig:CosParamW} show a comparison between our new Planck CMB constraints and those provided by the Planck 2015 public MCMC chains, respectively for the flat ${\Lambda}$CDM and wCDM models. In the first case, as expected, the new constraints on $\tau$ results in somewhat narrower credibility intervals and a lower value for $\sigma_8$ (by roughly 0.5$\sigma$). However, this also impacts the other parameters to a similar amount with higher values of $\Omega_m$ and $\Omega_b$ and, correspondingly, a lower value of $H_0$. This latter change actually compensates in part the decrease in $\sigma_8$ so that, in the end, the net impact on late time structures and the cluster density is negligible (see predictions in \autoref{dndz}). In the $w$CDM case, the errors from the primary CMB alone are much larger and the shifts due to the lower value of $\tau$ are not as significant. Our updated Planck 2015 results still show good consistency with a $\Lambda$CDM model, with best-fitting parameters in slight tension with some observations of the late time large-scale structures \citep{hildebrandt17} or distance scale indicators like Cepheids \citep{riess18}. Mean and standard deviations for each parameter in our chains are provided in \autoref{append:CosPar}.

As the present paper was being submitted, the Planck Collaboration released their final set of cosmological parameters \citep{planck2018VI}, including improved data analysis, likelihoods, and the new constraints on optical depth and accurate polarisation power spectra at all scales. For reference, we incorporated these new constraints in the cosmological parameter tables provided in \autoref{append:CosPar}. Our results compare very well with the final Planck measurements. The uncertainties on individual parameters only decrease by 10--20\%\ and 20--30\% respectively for the $\mathrm{\Lambda}$CDM and wCDM models with the final results. In addition, the offset in the best-fit cosmological models is in all cases smaller than the final Planck uncertainties. Given the current constraining power of our XXL analysis, such differences would have negligible impact on the conclusions of our work. 

\section{Derived cosmological parameters}
\label{append:CosPar} 
This appendix lists all the cosmological parameter constraints obtained in this paper, together with similar 
constraints from the latest releases of the WMAP \citep{hinshaw13} and the Planck satellite \citep{planck2015XIII}.
In \autoref{tab:CosParamLambda} we provide results for the flat $\Lambda$CDM case, while \autoref{tab:CosParamW} 
shows the constraints achieved in a $w$CDM model. 

\begin{table*}
        \begin{center}
        \begin{tabular}{lcccccc}
        \hline\hline
        Parameter    & WMAP9  & Planck15 & Planck (this work) &  Planck18 & XXL-C1 & C1+KiDS\\
        \hline
        $\mathrm{h}$    & $0.700\pm0.022$   & $0.6783\pm0.0092$ & $0.6740\pm0.0069$  & $0.6736\pm0.0054$ & $0.609\pm0.073$ & $0.740\pm0.049$\\
        $\mathrm{\Omega_b}$     & $0.0463\pm0.0024$ & $0.0484\pm0.0010$ & $0.0489\pm0.0008$  & $0.0493\pm0.0006$ & $0.062\pm0.015$ & $0.042\pm0.007$\\
        $\mathrm{\Omega_m}$     & $0.279\pm0.025$   & $0.308\pm0.012$   & $0.313\pm0.009$    & $0.315\pm0.007$ & $0.399\pm0.094$ & $0.312\pm0.049$\\
        $\mathrm{\sigma_8}$     & $0.821\pm0.023$   & $0.8149\pm0.0093$ & $0.8108\pm0.0066$  & $0.8111\pm0.0061$ & $0.721\pm0.071$ & $0.719\pm0.064$\\
        $\mathrm{n_s}$          & $0.972\pm0.013$   & $0.9678\pm0.0060$ & $0.9651\pm0.0047$  & $0.9649\pm0.0042$ & $0.965\pm0.023$ & $1.07\pm0.13$\\
        $\mathrm{\tau}$         & $0.089\pm0.014$   & $0.066\pm0.016$   &  $0.0566\pm0.0083$ & $0.0543\pm0.0074$ & - & -\\
        \hline\hline
        \end{tabular}
        \caption{Primary CMB constraints for the flat $\mathrm{{\Lambda}CDM}$ model. The parameter value corresponds to the mean over the Markov chain, while the error shows the standard deviation.
        No results are provided for the combination of Planck and C1 clusters since,  for such a small number of parameters, the Planck constraints will fully dominate the results. 
        }
        \label{tab:CosParamLambda}
        \end{center}
\end{table*}

\begin{table*}
        \begin{center}
        \begin{tabular}{lccccc}
        \hline\hline
        Parameter    & Planck15 & Planck (this work)  &  Planck18 & XXL-C1 &  C1+Planck \\
        \hline
        $\mathrm{h}$             & $0.82\pm0.12$       & $0.83\pm0.11$     & $0.87\pm0.09$ & $0.669\pm0.070$ &  $0.681\pm0.065$  \\
        $\mathrm{\Omega_b}$      & $0.035\pm0.011$     & $0.0343\pm0.0097$ & $0.0310\pm0.0071$ & $0.051\pm0.011$ &  $0.0491\pm0.0090$   \\
        $\mathrm{\Omega_m}$      & $0.224\pm0.074$     & $0.219\pm0.063$   & $0.197\pm0.046$ & $0.328\pm0.067$ &  $0.316\pm0.060$   \\
        $\mathrm{w}$         & $-1.41\pm0.35\ \ \,$      & $-1.44\pm0.30\ \ \,$    & $-1.57\pm0.25$\ \ \, &  $-1.53\pm0.62\ \ \,$  &  $-1.02\pm0.20\ \ \,$   \\
        $\mathrm{\sigma_8}$      & $0.925\pm0.094$     & $0.930\pm0.082$   & $0.964\pm0.069$ &  $0.775\pm0.078$ &  $ 0.814\pm0.054$  \\
        $\mathrm{n_s}$           & $0.9681\pm0.0061$   & $0.9669\pm0.0048$ & $0.9666\pm0.0041$ & $0.966\pm0.023$ &  $ 0.9649\pm0.0048$  \\
        $\mathrm{\tau}$          & $0.060\pm0.019$     & $0.055\pm0.009$   & $0.052\pm0.007$ &  -               & $0.0559\pm0.0087$ \\
        \hline\hline
        \end{tabular}
        \caption{Primary CMB constraints for the flat $\mathrm{wCDM}$ model.
        The parameter value corresponds to the mean over the Markov chain, while the error shows the standard deviation.}
        \label{tab:CosParamW}
        \end{center}
\end{table*}

\section{Impact of priors on the XXL C1 analysis}
\label{append:PriorImpact}
In \autoref{sec:Methods},  we describe a number of priors applied to the analysis of the XXL clusters alone in order to fix some parameters that our clusters cannot efficiently constrain ($n_s$, $\Omega_b$ through the combination $\Omega_b h^2$) and mitigate the degeneracy between $h$, $\Omega_m$, and $\sigma_8$ (using priors on $\Omega_m h^2$ and $h$ separately). Of course, these priors could have a significant impact on the comparison between the XXL clusters and Planck.  

In order to assess the importance of our choice of priors, we used importance sampling methods to modify the priors on our chains and derive alternative constraints:
\begin{itemize}
   \item \underline{Impact of Planck derived priors}. Our priors on $n_s$, $\Omega_b h^2$, and $\Omega_m h^2$ are centred on the Planck best-fit value, with Gaussian errors scaled by a factor 5 with respect to the Planck constraints. We opted for a factor of 5 in order not to force the XXL C1 constraints toward an  artificial agreement with Planck, but other choices were possible. In \autoref{tab:PriorLCDM} (for the ${\Lambda}$CDM model) and \ref{tab:PriorWCDM} (for the $w$CDM model), we present alternative constraints rescaling instead the errors by factors of 10, 3, and 1. The results are essentially the same with slight but insignificant shifts of the average values for all parameters. The resulting errors on $n_s$ directly scale with the width of the priors, as expected since the XXL C1 cluster alone do not bring significant constraints on this parameter. For all other parameters, the errors do not change significantly. 
   \item We also performed a similar exercise for the prior on $h$. For priors still centred on $h=0.7$, we changed the Gaussian standard deviation from the initial 0.1 to 0.05 and 0.2. The XXL clusters alone favour a value of $h$ lower than 0.7 in for both cosmological models, and therefore tighter priors on $h$ push the best-fit value higher. Given the Planck priors on $\Omega_b h^2$ and  $\Omega_m h^2$, which the cluster fit tightly follow, the values of the matter densities diminish accordingly. Shifts on $\sigma_8$ and $w$ also occur, but stay well within 1$\sigma$.
\end{itemize}

From these basic sanity checks, we conclude that the results presented in the paper for the XXL C1 clusters alone do not depend much on the details of our chosen priors and can be considered robust.

\begin{table*}
        \begin{center}
        \begin{tabular}{cccccccc}
        \hline\hline
    Parameter        &      Default       &  $\sigma(\mathrm{Planck})\times 10$  &   $\sigma(\mathrm{Planck})\times 3$    &    $\sigma(\mathrm{Planck})\times 1$    &      $\sigma(h)=0.2$      &    $\sigma(h)=0.05$  \\
        \hline
    $\mathrm{h}$ & $0.609\pm0.073$  & $0.615\pm0.075$  &  $0.608\pm0.073$  &  $0.608\pm0.073$  & $0.572\pm0.075$  &  $0.668\pm0.048$  \\
    $\mathrm{\Omega_b}$  &  $0.062\pm0.015$  &  $0.061\pm0.015$ & $0.063\pm0.015$  & $0.063\pm0.015$  & $0.071\pm0.017$  & $0.051\pm0.008$  \\
    $\mathrm{\Omega_m}$ & $0.399\pm0.094$ & $0.395\pm0.093$  & $0.401\pm0.095$  & $0.401\pm0.096$ & $0.452\pm0.106$ & $0.326\pm0.049$ \\
    $\mathrm{\sigma_8}$ & $0.721\pm0.071$  & $0.716\pm0.076$ & $0.720\pm0.071$ & $0.721\pm0.070$ & $0.706\pm0.073$ & $0.744\pm0.065$  \\
    $\mathrm{n_s}$  & $0.965\pm0.023$ & $0.964\pm0.041$ &  $0.965\pm0.014$ & $0.965\pm0.005$ & $0.964\pm0.023$ &  $0.966\pm0.023$ \\
\hline\hline
        \end{tabular}
        \caption{Impact of priors on the XXL C1 cosmological fits for the $\Lambda$CDM model. 
        The XXL derived constraints are provided for different widths of the priors on $n_s$, $\Omega_b h^2$, 
        and $\Omega_m h^2$, rescaling the Planck constraints by a factor of 10, 3, and 1 instead of 
        the factor 5 used for the main results.}
        \label{tab:PriorLCDM}
        \end{center}
\end{table*}

\begin{table*}
        \begin{center}
        \begin{tabular}{ccccccc}
        \hline\hline
    Parameter        &      Default       &  $\sigma(\mathrm{Planck})\times 10$  &   $\sigma(\mathrm{Planck})\times 3$    &    $\sigma(\mathrm{Planck})\times 1$    &      $\sigma(h)=0.2$      &    $\sigma(h)=0.05$  \\
        \hline
    $\mathrm{h}$               &   $0.669\pm0.070$  &   $0.659\pm0.071$  &   $0.670\pm0.069$  & $0.665\pm0.069$  &   $0.660\pm0.079$  & $0.689\pm0.047$  \\
    $\mathrm{\Omega_b}$        &   $0.051\pm0.011$  &   $0.053\pm0.012$ &  $0.051\pm0.011$  & $0.052\pm0.011$  &   $0.053\pm0.012$  & $0.047\pm0.007$  \\
    $\mathrm{\Omega_m}$  & $0.328\pm0.067$  & $0.335\pm0.067$  & $0.327\pm0.067$  & $0.332\pm0.068$  & $0.338\pm0.075$  &  $0.304\pm0.044$  \\
    $\mathrm{w}$ &  $-1.531\pm0.621\ \ \,$  &  $-1.587\pm0.606\ \ \,$  &  $-1.509\pm0.626\ \ \,$  & $-1.484\pm0.597\ \ \,$  & $-1.508\pm0.634\ \ \,$  & $-1.574\pm0.592\ \ \,$  \\
    $\mathrm{\sigma_8}$ & $0.775\pm0.078$  & $0.776\pm0.075$ & $0.774\pm0.079$  & $0.770\pm0.079$ & $0.771\pm0.080$  &   $0.787\pm0.075$  \\
    $\mathrm{n_s}$  &  $0.966\pm0.023$ &  $0.972\pm0.044$  &   $0.965\pm0.014$  & $0.965\pm0.005$  & $0.965\pm0.023$  &   $0.966\pm0.023$  \\
    \hline\hline
        \end{tabular}
        \caption{Impact of priors on the XXL C1 cosmological fits for the $w$CDM model. 
        The XXL derived constraints are provided for different widths of the priors on $n_s$, $\Omega_b h^2$, 
        and $\Omega_m h^2$, rescaling the Planck constraints by a factor of 10, 3, and 1 instead of 
        the factor of 5 used for the main results.}
        \label{tab:PriorWCDM}
        \end{center}
\end{table*}

\section{Quantifying the consistency of different probes}
\label{append:IOI}
To quantitatively assess the compatibility of our XXL C1 results with the Planck constraints, we rely on the Index of Inconsistency \citep[IOI,][]{lin17}.
To compare two datasets given a model, it simply measures the multi-dimensional distance between the best fits for each probe, $\bm{\mu}=\bm{\mathcal{P}^{(1)}} - \bm{\mathcal{P}^{(2)}}$, using the covariance of each fit ($\bm{C^{(1)}}$, $\bm{C^{(2)}}$) to define a metric as 
\begin{equation}
    \mathrm{IOI} = \frac{1}{2}\bm{\mu}^T\left(\bm{C^{(1)}}+\bm{C^{(2)}}\right)^{-1}\bm{\mu}.
\end{equation}

The interpretation of the IOI by \citet{lin17} relies on assigning compatibility levels for different ranges of the parameter, in a similar manner to the Jeffreys scale \citep{jeffreys61} used for model selection in Bayesian statistics.
However, the justification for this procedure remains rather vague and, strangely, does not depend on the number of parameters in the model, $N_p$. 
More interestingly, the authors correctly note the functional similarity of this statistic to $\chi^2$ and deduce that the confidence level can be derived as $n$-$\sigma = \sqrt{2 \mathrm{IOI}}$ when comparing two one-dimensional distributions. 
Actually, for posteriors approaching Gaussian distributions, we can show that $2\mathrm{IOI}$ should be distributed as a $\chi^2$ distribution with $N_p$ degrees of freedom.

In our case, since our posteriors deviates slightly from Gaussian distributions, we prefer to connect the measured $\mathrm{IOI}$ with confidence levels using Monte Carlo simulations. To do so, we translate the posterior distributions by substracting from them the best-fit parameters. The two posteriors are therefore centred on the same value and the points in our chain represent random fluctuations due to the precision of each experiment when both originate from the same model parameters. We use these fluctuations to generate draws of $\bm{\mu}$ and  the corresponding $\mathrm{IOI,}$ and to obtain a cumulative probability distribution for the $\mathrm{IOI}$. Finally, we estimate from this the probability to exceed (PTE) the observed $\mathrm{IOI}$ and the corresponding significance level. Since our Planck and XXL C1 posterior are not too different from Gaussian distributions, the significance levels obtained by this method are very similar to the values obtained from the identification with a $\chi^2$.

\section{Notations for galaxy cluster quantities}
\label{Append:Notations}
Throughout the paper we use a consistent set of notations laid out for the entire XXL survey to designate cluster physical quantities. Subscripts indicate the extraction radius within which the value was measured  or, when the quantity is the radius itself, its definition. A unitless extraction radius, usually 500, refers to an overdensity factor with respect to the critical density of the Universe. When relevant, an additional flag may be appended to the radius definition to specify the origin of the overdensity radius estimate, \texttt{WL} for a direct weak lensing mass estimate or \texttt{MT} when it relies on the measured X-ray temperature combined with a scaling relation. Finally, an \texttt{XXL} superscript for a luminosity explicitly indicates that it was estimated for the rest frame [0.5--2] keV band and corrected for galactic absorption. Subscripts and superscripts may be omitted when referring to generic quantities, for which the exact definition is irrelevant.

\end{appendix}

\end{document}